\documentclass[
reprint,
superscriptaddress,
amsmath,amssymb,
prl,
nolongbibliography
]{revtex4-2}
 
\usepackage{mathtools}
\usepackage{graphicx}
\usepackage[colorlinks=true,linkcolor=blue,citecolor=blue,urlcolor=blue,pdfauthor={ },pdftitle={ },pdfsubject={ },pdfkeywords={ }]{hyperref}
\usepackage{times}
\usepackage[qm]{qcircuit}
\usepackage{braket}
\usepackage{enumitem}
\usepackage{dcolumn}
\usepackage{bm}

\usepackage{siunitx}
\usepackage{upgreek}

\usepackage{color}
\usepackage{tikz}
\usepackage{pgf}
\usetikzlibrary{spy}

\definecolor{dred}{rgb}{.8,0.2,.2}
\definecolor{ddred}{rgb}{.8,0.5,.5}
\definecolor{dblue}{rgb}{.2,0.2,.8}
\definecolor{dgreen}{rgb}{.2,0.5,.2}

\newcommand*{\qcircuit}{$\begin{matrix}\Qcircuit @C=2em @R=2em}
\newcommand*{\eqcircuit}{\end{matrix}$}

\DeclareMathOperator{\Tr}{Tr}

\graphicspath{{figures/}}

\newcommand*{\physus}{Department of Physics, State Key Laboratory of Quantum Functional Materials, and Guangdong Basic Research Center of Excellence for Quantum Science, Southern University of Science and Technology, Shenzhen 518055, China}

\newcommand*{\qsc}{Quantum Science Center of Guangdong-HongKong-Macao Greater Bay Area, Shenzhen 518045, China}
\newcommand*{\sppeqnu}{ School of Physics and Physical Engineering, Shandong Provincial Key Laboratory of Laser Polarization and Information Technology, Qufu Normal University, Qufu 273165, China}

\begin{document}

\preprint{APS/123-QED}

\title{Reversing Heat Flow by Coherence in a Multipartite Quantum System}

\author{Keyi Huang}
\thanks{These authors contributed equally to this work.}
\affiliation{\physus}

\author{Qi Zhang}
\thanks{These authors contributed equally to this work.}
\affiliation{\sppeqnu}

\author{Xiangjing Liu}
\thanks{These authors contributed equally to this work.}
\affiliation{Nanyang Quantum Hub, School of Physical and Mathematical Sciences, Nanyang Technological University, Singapore}
\affiliation{Centre for Quantum Technologies, Nanyang Technological University, 637371, Singapore}

\author{Ruiqing Li}
\affiliation{\physus}

\author{Xinyue Long}
\affiliation{\qsc}

\author{Hongfeng Liu}
\affiliation{\physus}

\author{Xiangyu Wang}
\affiliation{\physus}

\author{Yu-ang Fan}
\affiliation{\physus}

\author{Yuxuan Zheng}
\affiliation{\physus}

\author{Yufang Feng}
\affiliation{\physus}

\author{Yu Zhou}
\affiliation{Ministry of Industry and Information Technology Key Lab of Micro-Nano Optoelectronic Information System, Guangdong Provincial Key Laboratory of Semiconductor Optoelectronic Materials and Intelligent Photonic Systems, Harbin Institute of Technology, Shenzhen 518055, China}
\affiliation{\qsc}

\author{Jack Ng}
\affiliation{\physus}

\author{Xinfang Nie}
\affiliation{\qsc}
\affiliation{\physus}

\author{Zhong-Xiao Man}
\email{zxman@qfnu.edu.cn}
\affiliation{\sppeqnu}

\author{Dawei Lu}
\email{ludw@sustech.edu.cn}
\affiliation{\physus}
\affiliation{\qsc}

\date{\today}

\begin{abstract}
The second law of thermodynamics dictates that heat flows spontaneously from a high-temperature entity to a lower-temperature one. Yet, recent advances have demonstrated that quantum correlations between a system and its thermal environment can induce a reversal of heat flow, challenging classical thermodynamic expectations. Here, we experimentally demonstrate that internal quantum coherence in a multipartite spin system can also reverse heat flow, without relying on correlations with the environment. Under the collision model with cascade interaction, we verify that both the strength and the phase of the coherence term determine the direction and magnitude of energy transfer.These results enable precise control of heat flow using only local quantum properties.
\end{abstract}

\maketitle

\emph{\bfseries Introduction.}---Controlling heat flow is a central task in thermodynamics, particularly in the context of cooling~\cite{sherwin1993introduction,doi:10.1126/science.1134008,Tan2017,PhysRevLett.129.100603,felce2020quantum,PhysRevLett.132.210403,aamir2025thermally,cao2022quantum}.
 Yet, the second law of thermodynamics imposes fundamental limitations. According to Clausius, heat flows spontaneously from a hot system to a cold one~\cite{clausius1879mechanical}.  Insights from information theory and quantum mechanics have further deepened our understanding of this foundational principle~\cite{parrondo2015thermodynamics,rio2011thermodynamic,PhysRevLett.100.080403,lieb2000fresh,lieb2002mathematical,sagawa2008second,PhysRevX.5.021001,dahlsten2011inadequacy,mmode2015limitations,Bera2017}. The Clausius statement, however, strictly applies only when the two systems are initially uncorrelated. When correlations are present, the direction of heat flow can be reversed, allowing heat to flow from the colder system to the hotter one. This phenomenon, known as anomalous heat flow, has been both theoretically predicted~\cite{jevtiv2012maximally,PhysRevE.77.021110,Bera2017,PhysRevE.81.061130} and experimentally verified~\cite{Micadei2019}.

However, realizing anomalous heat flow requires control over both the system and the environment, which is often unfeasible in realistic settings. Recent advances in quantum thermodynamics~\cite {RevModPhys.89.041003,PhysRevA.93.052335,meng2025quantum,liu2022thermodynamics,harshit2023re,elouard2017role,elouard2017alexia,huang2023engines,watanabe2025universal,lumbreras2025quantum}, especially the thermodynamic role of coherence~\cite{PhysRevX.5.021001,korzekwa2016extraction,mmode2015limitations,PRXQuantum.3.040323}, have suggested a way to bypass this limitation. The first law of energy conservation imposes restrictions on the evolution of quantum coherence: it separates the evolution of coherences within the same energy levels from that between different energy levels~\cite{PhysRevLett.120.150602}. As a result, coherence within degenerate energy levels can affect energy occupations, making it possible to control heat flow without access to the environment.

In this work, we experimentally investigate the influence of quantum coherence in multipartite spin system on heat transport. We prepare quantum states of a molecule system with controllable coherence using nuclear magnetic resonance (NMR) technology~\cite{Levitt2013SpinDynamics,PhysRevLett.133.140602,Fan2023,PhysRevLett.126.110502,PhysRevA.108.062211,PhysRevLett.108.040401,PhysRevLett.115.120403,PhysRevA.91.022121,zheng2025exp} and expose them to a thermal bath within a collision model featuring cascade interactions. As a result, we observe that coherence, particularly the phase of the coherence term, significantly influences the direction of heat flow~\cite{soldati2024coolinglimitscoherentrefrigerators}. Such behavior appears to violate the second law of thermodynamics, as heat can flow from cold bath to hot system. To understand this apparent contradiction, we further analyze the thermodynamic behavior of the system and demonstrate that the reversal of heat flow is driven by the consumption of internal coherence within the system, which can be reflected in its free energy. In addition, the state alone does not determine the direction of heat flow in our setting. We thus employ the concept of \emph{apparent temperature} (AT), an effective temperature depending on both the system’s state and dynamics, to capture the influence of coherence on energy exchange~\cite{Latune_2019}. We confirm that heat flows from regions with higher AT to those with lower AT. Based on these results, we demonstrate precise control of heat flow by manipulating coherence within degenerate energy levels of the system. Furthermore, we vary the interaction order in the collision model to study its impact on heat transfer, revealing additional degrees of freedom for heat regulation.

\begin{figure}[htbp]
\centering
\includegraphics[width=0.9\linewidth]{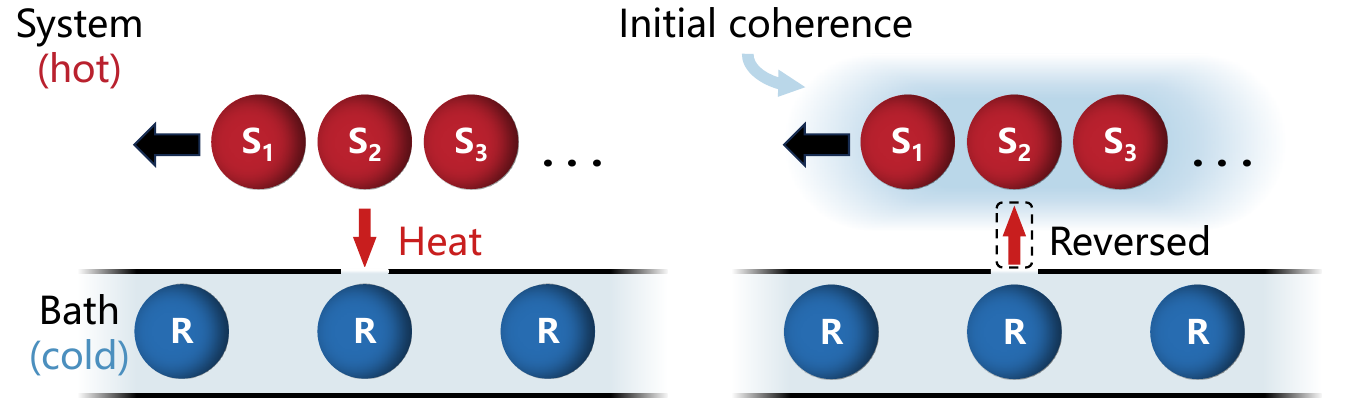}
\caption{ Collision model with cascade interaction. The system and the bath are uncorrelated. The system consists of multiple subsystems and is assumed to be in a thermal state at a higher temperature, while the thermal bath comprises an infinite number of identical particles at a lower temperature. In this collision model, each subsystem sequentially interacts with a bath particle in a fixed order, such as $S_1$–$S_2$–$S_3\cdots$. In this scenario, heat flows from the higher-temperature subsystems to the lower-temperature bath. However, injecting coherence into the system can modify this behavior and even reverse the direction of heat flow.}
\label{Fig1}
\end{figure}

\emph{\bfseries Framework.}---We formulate a collision model with cascade interaction to explore the influence of system coherence on heat transfer. The system is initially uncorrelated with the heat bath and comprises $n$ two-level spin subsystems, labeled as $S_1, S_2, \cdots, S_n$, respectively. The Hamiltonian of each spin is given by $H_{k} = \frac{\delta}{2} \sigma_z^k$, where $\delta$ denotes the energy gap and $\sigma_z^k$ is the Pauli-$z$ operator of the $k$-th spin. The heat bath is composed of an infinite sequence of identical thermal particles $R$, each in the state $\rho_R = e^{-\beta_R H_R} / \text{Tr} (e^{-\beta_R H_R})  $ where the inverse temperature $\beta_R= \frac{1}{T_R}$ and $H_R = \frac{\delta}{2} \sigma_z^R$.
The system is then brought into contact with the heat bath in a cascaded manner: the first subsystem $S_1$ interacts with $R$, followed by $S_2$, then $S_3$, and so on,  until $S_n$. Each interaction is governed by $ U_k = e^{-i\tau H_{kR}}$, with the interaction time $\tau$, and the interaction Hamiltonian $H_{kR}= g(\sigma_+^k \sigma_-^R + \sigma_-^k \sigma_+^R)$. Here, $\sigma_\pm^k$ and $\sigma_\pm^R$ are the standard ladder operators for the subsystem and bath particle, respectively, and $g$ denotes the interaction strength. After the bath particle $R$ interacts with all $n$ subsystems, it is replaced by a new one.  The interaction also satisfies the first law of energy conservation $[U_{k}, H_k + H_R] = 0$. Therefore, all energy changes in each subsystem can be clearly attributed to energy flowing to or from the bath particle.

We next consider the impact of coherence on heat transfer. Suppose that the subsystems are initially prepared in thermal states with temperatures exceeding $T_R$, then heat flows from the system to the heat bath according to the second law. We subsequently inject coherence into the system to investigate its effect on heat transfer. A schematic illustration is given in Fig.~\ref{Fig1}. In the Supplemental Information (SI)~\cite{supply}, we show that the energy conservation condition ensures that only coherence between degenerate energy levels of the system can influence its energy occupation.
Therefore, we inject coherence into the degenerate energy levels such that the initial state $\rho_S^0$ is of the form
\begin{equation}    
    \rho_S^0 = \rho_S^\mathrm{th} + \sum_{p<q} (c_{pq} \chi_{pq} + \text{h.c.}),
\end{equation}
where $\rho_S^\mathrm{th} = \bigotimes_{k=1}^n \rho_{k}^\mathrm{th}$ is the product of local thermal states $\rho_{k}^\mathrm{th} = e^{-\beta_k H_{k}} /  \mathrm{Tr}(e^{-\beta_k H_{k}})$ with $\beta_k=\frac{1}{T_k} \leq \beta_R$. The coefficients $c_{pq} = \lambda e^{i\alpha}$ characterize the coherence between subsystems, where $\lambda$ and $\alpha$ denote the strength and phase, respectively. These parameters govern the influence of coherence on heat flow. The terms $\chi_{pq}$ are constructed to introduce off-diagonal coherence between degenerate energy levels $\left|0_p 1_q\right>$ and $\left|1_p 0_q\right>$ for subsystems $S_p$ and $S_q$, without altering any local reduced state or the internal energy of the system, with details in SI~\cite{supply}. Based on this initial state and the collision model, we experimentally demonstrate that coherence present in the system modifies heat flow and enables its precise regulation.

\begin{figure}[htbp]
\centering
\includegraphics[width=1\linewidth]{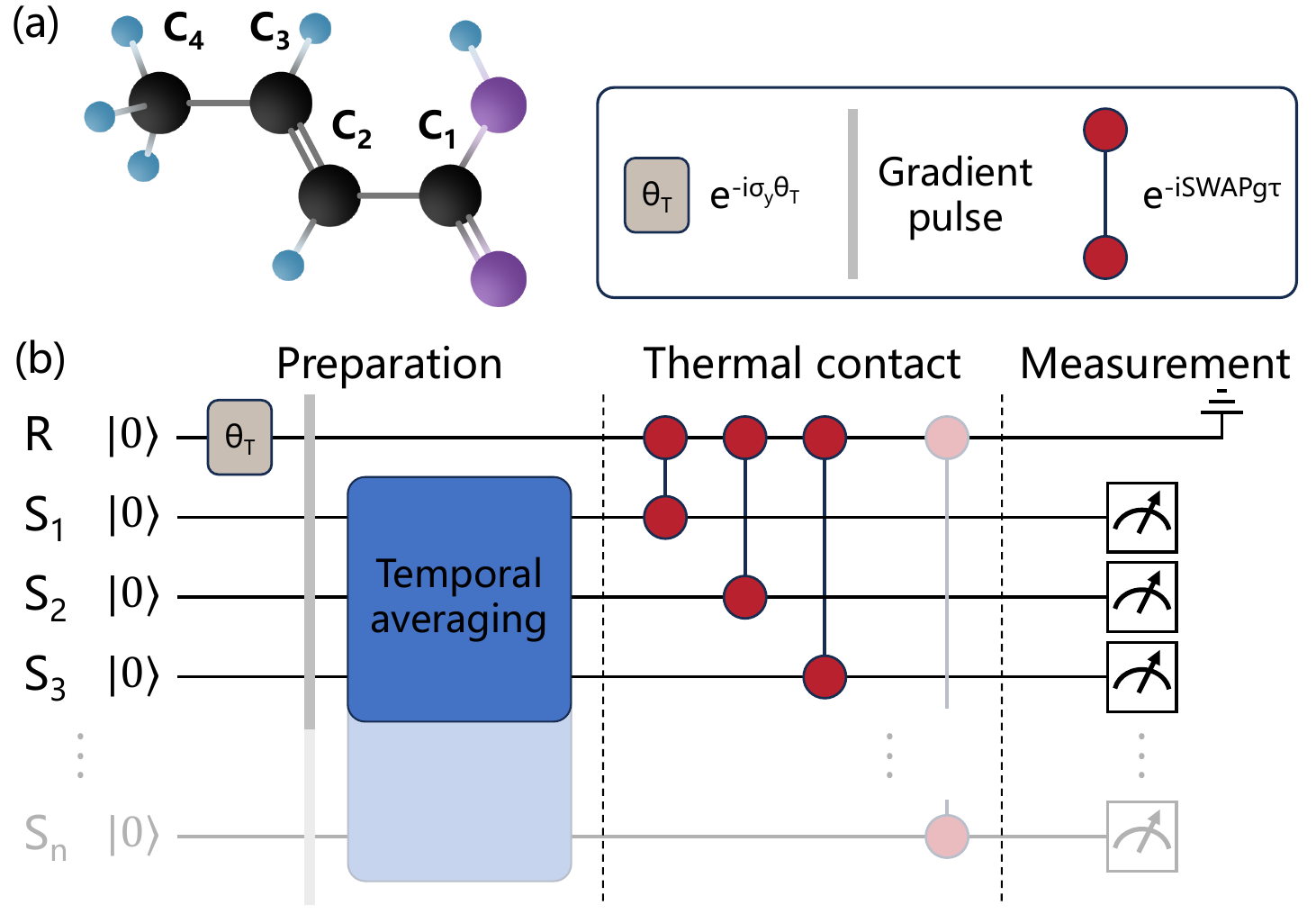}
\caption{(a) Molecular structure of crotonic acid used as the four-qubit NMR system. The nuclear spins of C$_1$, C$_2$, C$_3$, and C$_4$ are designated as subsystems S$_1$, S$_2$, S$_3$, and bath particle R, respectively. (b) Quantum circuit implementing the collision model with cascade interaction. It consists of three segments corresponding to the stages outlined in the main text. The first two operations initialize the bath particle. The system state with controlled coherence is prepared using temporal averaging, which involves applying different operations at distinct times. The thermal contact stage includes $n$ partial SWAP operations, with $n=3$ in our experiment.}
\label{Fig2}
\end{figure}

\emph{\bfseries Setup.}---The experiments are performed on a Bruker 300 MHz spectrometer at room temperature~\cite{Baugh2005,PhysRevLett.100.140501}. We use a $^{13}$C-labeled trans-crotonic acid sample dissolved in $d_6$ acetone to observe heat flow between two parts of the molecule~\cite{PhysRevLett.134.040201,PhysRevLett.129.070502}. The molecular structure of the sample is shown in Fig.~\ref{Fig2}(a). Here, we present the experimental setup for the case of $n=3$, where three $^{13}$C nuclei serve as the system consisting of three two-level spins and the other one plays the role of the heat bath particle. The internal Hamiltonian of the system is given by
\begin{equation}
\mathcal{H}_{\text{NMR}}=-\sum_k\frac{\omega_k}{2} \sigma_z^k + \frac{\pi}{2} \sum_{p<q} J_{pq} \sigma_z^p \sigma_z^q,    
\end{equation}
where $\omega_k$ is the Larmor frequency, and $J_{pq}$ denotes the coupling strength between the $p$-th and $q$-th spins. Specific values for $\omega_k$ and $J_{pq}$ can be found in SI~\cite{supply}. Transverse radio-frequency pulses are applied to execute single-qubit rotations, and precise control over nuclear spin couplings allows us to realize the desired interactions.

\emph{\bfseries Procedure.}---In our experiment, the contact between the system and an infinite-particle heat bath is emulated by repeating similar interaction processes, as described in the SI~\cite{supply}. Each repetition consists of three stages [Fig.~\ref{Fig2}(b)]: (i) preparation of the initial state, (ii) thermal contact following the collision model, and (iii) extraction of heat flow from energy measurements.

\begin{figure*}[!htbp]
\centering
\includegraphics[width=1\linewidth]{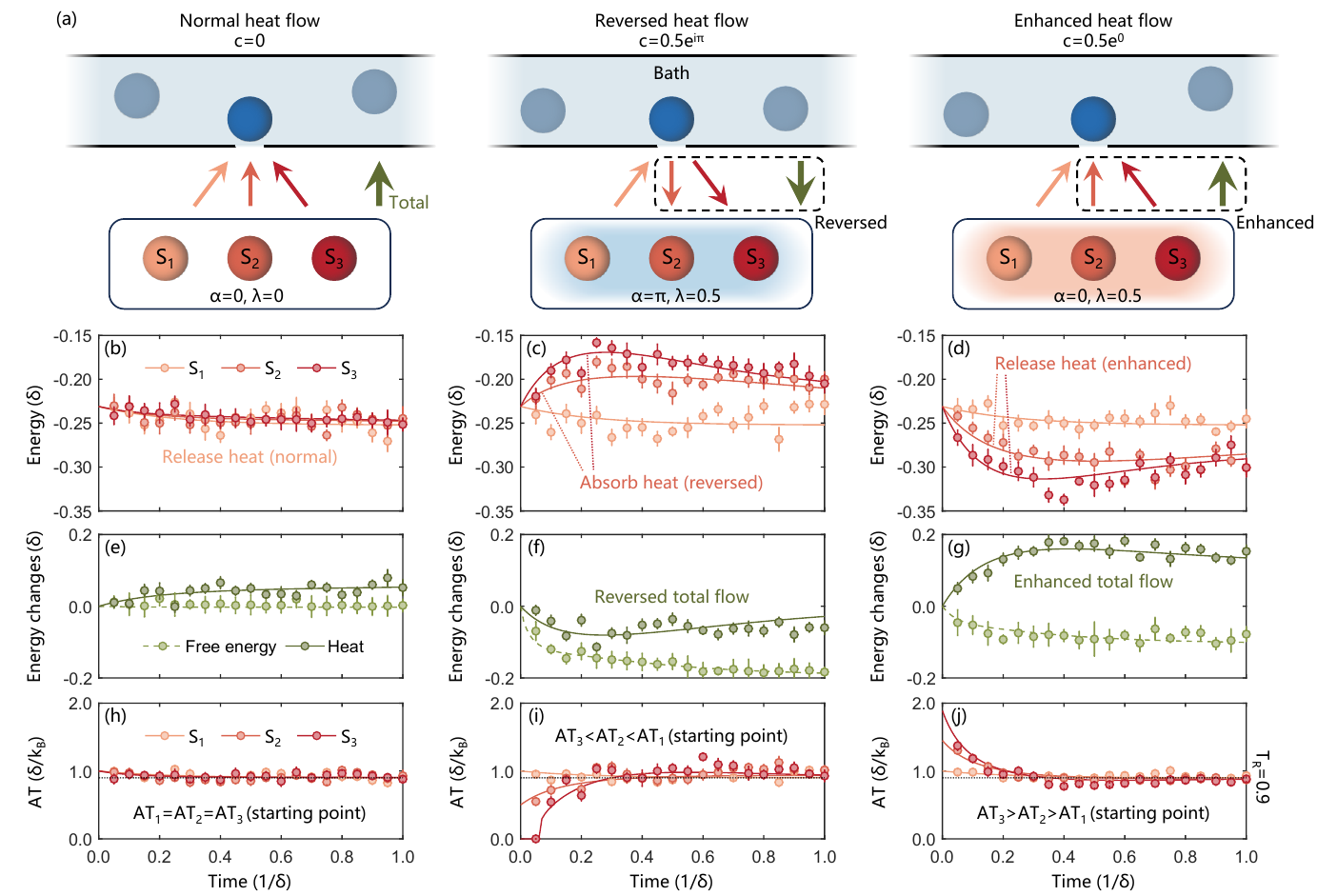}
\caption{(a) Overview of energy transfer dynamics in the thermal contact. (b)-(d) Time evolution of the energy of each subsystem for different initial coherence values. The energy change in $S_3$ is consistently larger than that in $S_2$, while $S_1$ follows the behavior of standard thermal contact. (e)-(g) Heat transfer and free energy changes in the system. Heat flowing from system to bath is defined as positive, with the system prepared at higher local temperature. With coherence $c = 0.5e^{i\pi}$, heat tends to flow from the bath to the system; in contrast, for the same coherence amplitude but with phase $\alpha = 0$, heat flows from the system to the bath. (h)-(j) ATs of $S_1$, $S_2$, and $S_3$ as functions of interaction time, corresponding to the experimental settings in (b)-(d). The dashed line indicates the temperature of the bath, which is equal to its AT. Symbols denote experimental data; lines represent numerical simulations.}
\label{Fig3}
\end{figure*}

(i) The initial state preparation involves both the system and the bath components in the collision model. The bath particle is prepared in a thermal state by adjusting population distributions through single-qubit rotations, followed by a 1 ms \textit{z}-gradient pulse to eliminate residual coherence. The system is initialized in a mixed state with tunable coherence, constructed using temporal averaging techniques to independently control populations and coherence components~\cite{PhysRevA.57.3348}.

(ii) The collision model with cascade interaction is implemented through a sequence of unitary operations, each describing the interaction between a subsystem and a bath particle. These operations are applied sequentially according to the prescribed interaction order. At the operational level, each interaction is effectively implemented as a partial SWAP operation. In our experiment, this is realized by applying shaped radio-frequency pulses for a controlled short duration. The full sequence of three such interactions is executed in 40~ms, with a simulation fidelity over 99.5\%.

(iii) In the final stage, the heat flow is determined from changes in the system’s internal energy, based on energy conservation and the absence of external work~\cite{PhysRevX.5.031044,PhysRevLett.132.210403}. The internal energy is obtained by reconstructing the system state at different times during the cascade process and computing its expectation value with respect to the system Hamiltonian. By comparing these values before and after thermal contact, we extract the net energy exchange, which corresponds to the heat flow. Further details are provided in SI~\cite{supply}.

\emph{\bfseries Experimental result.}---We begin by studying how quantum coherence affects heat transfer in a three-spin system within the collision model featuring cascade interactions. In this experiment, we consider a simple coherence structure in which the coherence terms between any two subsystems are set to be equal ($c_{pq}=c$). The thermal contact is implemented in the fixed order $S_1$-$S_2$-$S_3$, with an interaction time of $\tau = 0.01\delta^{-1}$, coupling strength $g = 20~\delta$, and initial temperatures of the subsystems set to $T_k = \delta/k_\mathrm{B}$. The bath temperature is set at $T_R = 0.9~\delta/k_\mathrm{B}$. An overview of the energy transfer dynamics is illustrated in Fig.~\ref{Fig3}(a), and the detailed energy evolution over time for each subsystem at different coherence values is shown in Fig.~\ref{Fig3}(b)-\ref{Fig3}(d).

The results reveal several distinctive features induced by coherence. As a reference, in the absence of coherence ($c = 0$), energy transfer follows conventional thermodynamic expectations: each hot subsystem gradually releases heat to the cold bath, as shown in Fig.~\ref{Fig3}(b). When coherence is present, both the magnitude and direction of heat flow can be influenced by the phase of the coherence term. For example, when the coherence term is set to $c = -0.5$, corresponding to an amplitude of $0.5$ and a phase of $\pi$, heat is transferred from the cold bath to the hot system, indicating a reversal of the natural thermodynamic direction. This is evidenced by subsystems $S_2$ and $S_3$ absorbing heat from the lower-temperature bath [Fig.~\ref{Fig3}(c)]. In contrast, when the phase is $0$ with the same amplitude ($\lambda = 0.5$), heat transfer is enhanced along the conventional direction with more heat released to the bath [Fig.~\ref{Fig3}(d)].

Moreover, the thermal response is not uniformly distributed across the system. Due to the sequential nature of the interactions in the collision model, the influence of coherence accumulates progressively. Consequently, subsystem $S_3$ exhibits more substantial energy variation than $S_2$, regardless of whether coherence enhances or reverses the heat flow. Subsystem $S_1$, due to its dynamics being unaffected by coherence, always exhibits normal heat flow.

These results highlight the power of coherence: by tuning its strength and phase, heat flow can not only be enhanced but also reversed, allowing refrigeration where heat flows from a cold to a hot region, seemingly violating the second law of thermodynamics.

\emph{\bfseries Thermodynamic interpretation and apparent temperature.}---Next, we provide a thermodynamic resource analysis to resolve the apparent violation. This is illustrated using the simplest two-subsystem model, where we consider an initial state with coherence distributed across the system. During the first interaction between subsystem $S_1$ and a bath particle $R$, part of this coherence is transformed into a correlation between $S_2$ and $R$~\cite{esposito2010entropy}. Subsequently, when $S_2$ interacts with the same bath particle, this correlation affects the direction of energy flow. Overall, the consumption of coherence drives the reversal of heat flow while simultaneously enhancing the correlations between the system and the bath particles. A more general analysis is detailed in the SI~\cite{supply,PhysRevLett.113.140401}.

From the energetic perspective, changes in coherence can be reflected by variations in the system’s free energy. Therefore, we monitor the free energy of the system and the heat transferred to the bath particles during the experiment, as shown in Fig.~\ref{Fig3}(e)-\ref{Fig3}(g). In all scenarios, the system's free energy decreases during thermal contact. When the heat flows in the conventional direction, the reduction of the system's free energy is negligible [Fig.~\ref{Fig3}(e)]. In contrast, when the heat flows are anomalous, the free energy of the system is significantly consumed to drive the reversal and enhancement of heat flow [Fig.~\ref{Fig3}(f) and (g)]. For all cases, the overall process remains consistent with the second law of thermodynamics.

To further characterize the local thermodynamic behavior of each subsystem, we employ the concept of AT~\cite{Latune_2019}, which serves as an effective indicator of thermal behavior in coherent, non-equilibrium systems. For a given subsystem $S_k$ undergoing cascade interaction, the AT is defined as~\cite{PhysRevA.108.062211}
\begin{equation}
    AT_{k} = \delta \left( \ln \frac{P_k^0 +\mathcal{C}_k }{P_k^1 + \mathcal{C}_k } \right)^{-1},
\end{equation}
where $P^{0(1)}_k$ represents the ground state (excited state) population of subsystem $S_k$, and $\mathcal{C}_k$ represents the contribution from coherence between degenerate energy levels of $S_k$ and its preceding subsystem $S_p$, with details in SI~\cite{supply}.

We track the AT of each subsystem over the course of thermal contact. When $c = 0$, no coherence is present, the AT remains equal to the initial physical temperature, as shown in Fig.~\ref{Fig3}(h). In contrast, for coherent initial states ($\lambda \ne 0$), the AT deviates: it decreases when the coherence phase is $\alpha = \pi$ [Fig.~\ref{Fig3}(i)] and increases when the phase is $\alpha = 0$ [Fig.~\ref{Fig3}(j)]. Meanwhile, the AT of the bath remains equal to its physical temperature in the absence of coherence.

By comparing the AT with the energy changes in each subsystem, we observe that a subsystem absorbs heat from the bath when its AT is lower than that of the bath, and releases heat when it is higher. This confirms that the AT governs the direction of heat flow. Although heat may appear to flow from a cold region to a hot one, it consistently flows from a subsystem with higher AT to one with lower AT. Thus, we can obtain the condition for the reversal of heat flow, which aligns well with our conclusions derived from the coherence-based analysis shown in SI~\cite{supply}.

\begin{figure}[!htbp]
\centering
\includegraphics[width=1\linewidth]{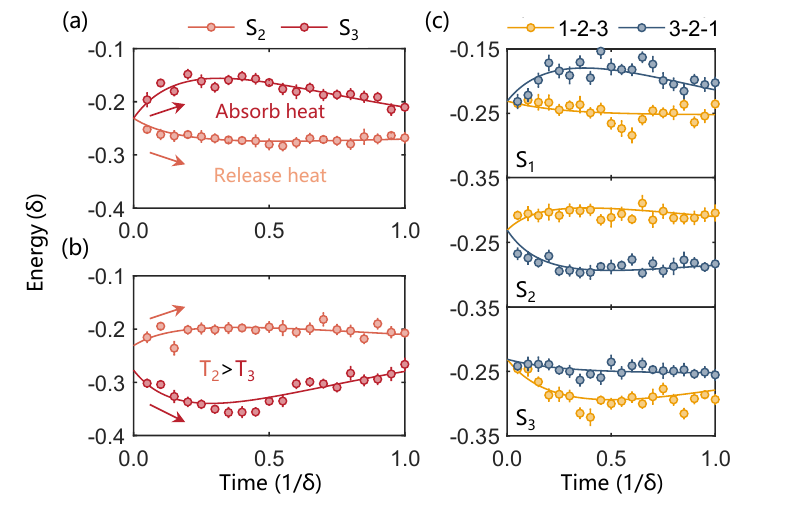}
\caption{(a) Heat flow behavior in $S_2$ and $S_3$ under equal initial temperatures. The coherence terms are set as $c_{12} = 0.3$, $c_{23} = c_{13} = -0.5$, leading to opposite heat flow directions in the two subsystems. (b) Reversed heat flow under a temperature gradient. The subsystem temperatures are $T_2 = \delta/k_{\mathrm{B}}$, $T_R= 0.9\delta/k_{\mathrm{B}}$ and $T_3 = 0.8\delta/k_{\mathrm{B}}$, with coherence set as $c_{12}=-0.5$ and $c_{23} = c_{13} = 0.5$. Despite the gradient, the hotter $S_2$ absorbs heat while the colder $S_3$ releases it. (c) Energy dynamics under different interaction sequences. The coherence is fixed at $c_{12} = -0.5$, $c_{13} = -0.1$, and $c_{23} = 0.5$. Changing the order of interaction among subsystems reverses the direction of heat flow in key ordinal positions, such as the second ($S_2$ for both sequence) and the final ($S_3$ for forward sequence and $S_1$ for reverse sequence) interacting spin. The other parameters are the same as in Fig.~\ref{Fig3}. }
\label{Fig4}
\end{figure}

\emph{\bfseries Heat flow regulation.}---Our results demonstrate that quantum coherence offers precise control over the direction and magnitude of heat flow in each subsystem. By tuning the initial coherence, we realize both enhancement and reversal of heat transfer under identical thermal conditions. In Fig.~\ref{Fig4}(a), all subsystems are initialized at the same temperature, yet coherence reverses the direction of heat flow between $S_3$ and the bath, while enhancing it at $S_2$. In a more extreme configuration [Fig.~\ref{Fig4}(b)], where $T_2 > T_R > T_3$, the hotter subsystem absorbs heat and the colder one releases it—demonstrating a clear departure from classical thermal expectations.

In addition, modifying the interaction sequence offers an alternative means to control heat flow, since the thermal behavior of each subsystem is shaped by the accumulated effects of preceding interactions. As shown in Fig.~\ref{Fig4}(c), switching the interaction order from $S_1$–$S_2$–$S_3$ to $S_3$–$S_2$–$S_1$, while keeping coherence fixed, reverses the direction of energy flow in some subsystems. For example, $S_2$ absorbs heat when it comes after $S_1$ in the sequence, but releases heat when it follows $S_3$. The subsystem that interacts with the bath last also exhibits distinct thermal responses: $S_3$ releases heat in the forward sequence, whereas $S_1$ absorbs heat in the reverse sequence. These results demonstrate that coherence and interaction order constitute two strategies for tuning quantum heat transport.

\emph{\bfseries Summary and discussion.}---Proper coherence can reverse the direction of heat flow. In particular, while Gibbs states are completely passive, coherence or correlations drive the system out of passivity and enable \textit{uphill} energy transfer in full accordance with the second law of thermodynamics~\cite{PhysRevE.91.052133,Sparaciari2017,Koukoulekidis2021geometryofpassivity}. The reversal of heat flow is determined by the AT gradient of the system and the bath, which embodies the role of coherence.  With the help of NMR technology, we prepared initial coherent mixed states and experimentally demonstrated how internal system coherence influences the direction and magnitude of heat flow.

Our findings reveal that both the strength and phase of coherence play key roles, with the phase being especially critical in determining the direction of heat transfer. Based on this, we demonstrated controllable heat flow regulation through coherence engineering. Furthermore, we investigated the role of interaction order in the collision model with cascade interaction and show that it can also serve as a mechanism for directing heat flow, independent of coherence changes. These results may inspire more possibilities for quantum cooling and thermodynamic computing~\cite{Huang_2025,conte2019thermodynamic}.

\emph{\bfseries Acknowledgement.}---We thank the anonymous reviewers for their valuable comments and suggestions that helped improve this work.
This work is supported by Pearl River Talent Recruitment Program (2019QN01X298), Guangdong Provincial Quantum Science Strategic Initiative (GDZX2303001, GDZX2200001, GDZX2403004), GuangDong Basic and Applied Basic Research Foundation (2022A1515110382), Shenzhen Fundamental Research Program (JCYJ202412023000152, JCYJ20241202123903005), National Natural Science Foundation of China Grants No. 12575024 and No. 12274257, and Shandong Provincial Natural Science Foundation (ZR2023LLZ015, ZR2024LLZ012). Xiangjing Liu is supported by the National Research Foundation through the NRF Investigatorship on Quantum-Enhanced Agents (Grant No. NRF-NRFI09-0010) and the National Quantum Office, hosted in A*STAR, under its Centre for Quantum Technologies Funding Initiative (S24Q2d0009), the Singapore Ministry of Education Tier 1 Grant RT4/23 and RG77/22 (S), the FQXi R-710-000-146-720 Grant “Are quantum agents more energetically efficient at making predictions?” from the Foundational Questions Institute, Fetzer Franklin Fund (a donor-advised fund of Silicon Valley Community Foundation).

\bibliography{bib}

\clearpage
\onecolumngrid 

\setcounter{equation}{0}
\setcounter{figure}{0}
\setcounter{table}{0}
\setcounter{section}{0}

\renewcommand{\theequation}{S\arabic{equation}}
\renewcommand{\theparentequation}{S\arabic{parentequation}}
\renewcommand{\thefigure}{S\arabic{figure}}
\renewcommand{\thetable}{S\arabic{table}}

\renewcommand{\theHequation}{supp.\arabic{equation}}
\renewcommand{\theHfigure}{supp.\arabic{figure}}
\renewcommand{\theHtable}{supp.\arabic{table}}
\renewcommand{\theHsection}{supp.\arabic{section}}

\begin{center}
\textbf{Supplemental Information for\\
Reversing Heat Flow by Coherence in a Multipartite Quantum System}
\end{center}

\section{Coherence Between Spins}

In the main text, we introduce a coherence term into the thermal state of every two spins. This coherence does not affect the local state of any individual spin. The total system state is given by  
\begin{equation}
\renewcommand{\theequation}{S\arabic{equation}}
    \rho_S^0 = \rho_S^\mathrm{th} + \sum_{p<q} \big( c_{pq} \, \chi_{pq} + \text{h.c.} \big),
\end{equation}
where $\rho_S^\mathrm{th} = \bigotimes_{k=1}^{n} \rho_k^\mathrm{th}$ denotes the thermal state of the entire system, expressed as a tensor product of the local thermal states $\rho_k^\mathrm{th}$. The second term represents the coherence, constructed from the coefficient $c_{pq} = \lambda e^{i\alpha}$, which encodes the coherence strength and phase, and the operator $\chi_{pq}$, which defines the coherence structure:  
\begin{equation}
\renewcommand{\theequation}{S\arabic{equation}}
    \chi_{pq} = \Big( \bigotimes_{k \neq p,q} \rho_k^{th} \Big)
    \otimes 
    \sqrt{P_p^0 P_p^1 P_q^0 P_q^1} \, |0_p1_q\rangle \langle 1_p0_q |.
\end{equation}
Here, $P_k^{0(1)}$ denotes the population in the ground (excited) state of the $k$-th spin. Importantly, the coherence defined in this form preserves the total system energy, as it commutes with the system Hamiltonian: $[H_S, \chi_{pq}] = 0$ with $H_S=\sum_{k}H_k$.

\section{Effect of Coherence Between Nondegenerate Levels} 

Consider the standard setting where the system $S$ and the environment $R$ are initially in thermal states with temperatures $T_S$ and $T_R$, respectively. Let us further assume that $T_S > T_R$. According to the second law of thermodynamics, heat flows from $S$ to $R$. Questions arise: what happens if we inject coherence into the system? Does coherence affect the direction of heat flow? Here, we show that coherence between nondegenerate levels does not affect heat flow.

The energy-preserving condition $[U, H_S + H_R] = 0$ imposes constraints on the evolution of quantum states. Assume that the Hamiltonian $H_S$ has no degenerate levels. Denote the set of all differences between the eigenfrequencies of $H_S$ as $\{\omega\}$. Then, the state of the system can be decomposed as
\begin{equation}
\renewcommand{\theequation}{S\arabic{equation}}
\rho_S = \sum_{\omega} \rho(\omega), \qquad
\rho(\omega) := \sum_{\substack{p,q \\ \omega_p - \omega_q = \omega}} \rho_{pq} |p\rangle \langle q|,
\end{equation}
where $H_S\ket{k} = \omega_k \ket{k}$ and the operators $\rho(\omega)$ are referred to as modes of nondegenerate coherence. These modes are characterized by the transformation property
\begin{equation}
\renewcommand{\theequation}{S\arabic{equation}}
    e^{-iH_S t} \rho(\omega) e^{iH_S t} = e^{-i\omega t} \rho(\omega).
\end{equation}

Under the assumption that the environment is initially in a thermal state $\gamma_R$ at temperature $T$, the effective channel acting on the system is given by
\begin{equation}
\renewcommand{\theequation}{S\arabic{equation}}
    \mathcal{E}_T(\cdot) = \Tr_R \left[ U (\cdot \otimes \gamma_R) U^\dagger \right].
\end{equation}
The energy-preserving condition implies that the channel $\mathcal{E}_T$ is symmetric under time translations, i.e.,
\begin{equation}
\renewcommand{\theequation}{S\arabic{equation}}
    e^{-iH_S t} \mathcal{E}_T(\rho_S) e^{iH_S t} = \mathcal{E}_T(e^{-iH_S t} \rho_S e^{iH_S t})
\end{equation}
for all $t$. This further implies that each mode of coherence evolves independently, such that
\begin{equation}
\renewcommand{\theequation}{S\arabic{equation}}
\mathcal{E}_T(\rho(\omega)) = \sigma(\omega)
\end{equation}
for all $\omega$. That is, each initial mode $\rho(\omega)$ is mapped to a corresponding mode $\sigma(\omega)$ of the final state via the thermal operation $\mathcal{E}_T$. Since heat transfer only involves the zero mode $\rho(0)$, coherence between nondegenerate levels does not influence it. Notably, coherence between degenerate levels is contained in the zero mode and can, in principle, affect the direction and magnitude of heat transfer, which is the central focus of our work.

\section{Experiment System}
\begin{figure}[htbp]
    \centering
    \includegraphics[width=1\linewidth]{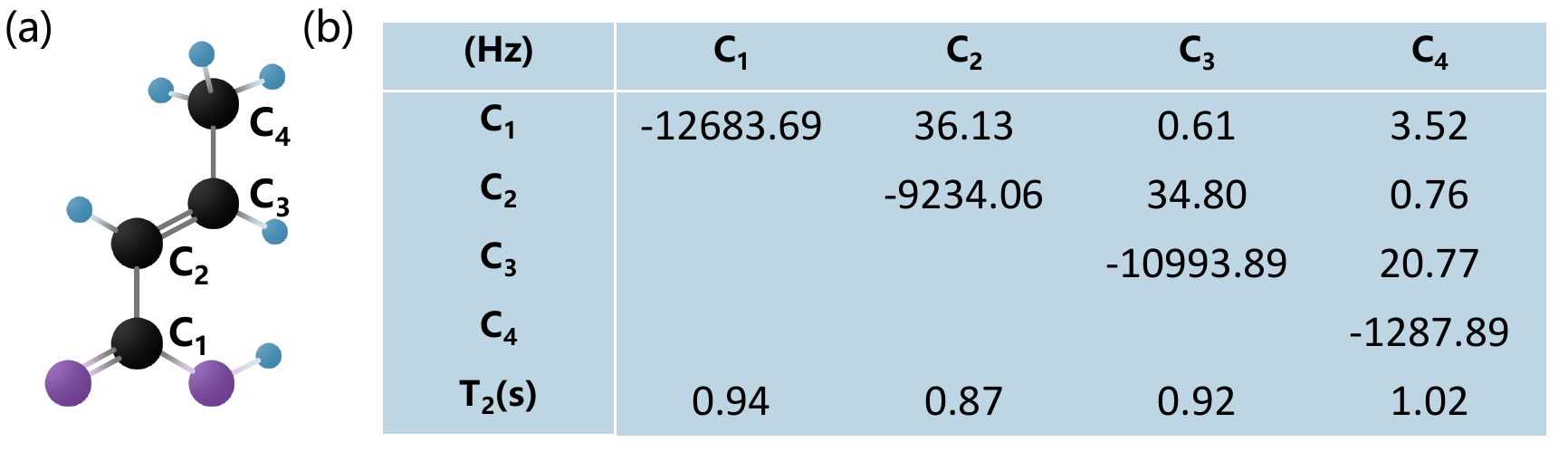}
    \caption{Structure and parameter.}
    \label{molecular}
\end{figure}

Our experiments are conducted on a nuclear magnetic resonance (NMR) quantum processor, using $^{13}$C-labeled crotonic acid molecules dissolved in d$_6$-acetone. The molecular structure is shown in Figure~\ref{molecular}(a). This molecule contains four coupled $^{13}$C nuclear spins, denoted as C$_1$, C$_2$, C$_3$, and C$_4$, which serve as subsystems S$_1$, S$_2$, S$_3$, and the bath particle R, respectively. All hydrogen atoms, including $\mathrm{H}_1$, $\mathrm{H}_2$, and the methyl group $\mathrm{M_H}$, are decoupled throughout all experiments.

The system is described by its internal Hamiltonian:
\begin{equation}
\renewcommand{\theequation}{S\arabic{equation}}
    \mathcal{H}_{\text{NMR}}=-\sum^4_{k=1}\frac{\omega_k}{2}\sigma^k_z+\frac{\pi}{2}\sum^4_{p<q}J_{pq}\sigma^p_z\sigma^q_z,
\end{equation}
where $\sigma_z$ is the Pauli matrix, and $\omega_k = 2\pi \nu_k$ is the Larmor frequency determined by the chemical shift $\nu_k$ of the $k$th spin. The coupling constants $J_{pq}$ represent the interaction strength between the $p$th and $q$th spins. These parameters are determined from the NMR spectrum measured at thermal equilibrium, and are listed in Figure~\ref{molecular}(b) with the Larmor frequencies and coupling strengths provided in the diagonal and off-diagonal positions, respectively. Based on the spin-spin couplings in the internal Hamiltonian, controlled two-spin interactions are realized via natural evolution under this Hamiltonian.

Control of individual spins is achieved by applying radio-frequency (rf) pulses, described by the Hamiltonian:
\begin{equation}
\renewcommand{\theequation}{S\arabic{equation}}
    H_{rf}=-\sum_k\gamma_kB_1[\cos{(\omega_{rf}t+\phi)} \sigma^k_x+\sin{(\omega_{rf}t+\phi)} \sigma^k_y],
\end{equation}
where $B_1$, $\phi$, and $\omega_{rf}$ denote the amplitude, phase, and frequency of the rf pulse, respectively, and $\gamma_k$ is the gyromagnetic ratio of the $k$th spin. Arbitrary single-spin operations are implemented by designing sequences of rf pulses with specific amplitudes and phases. Combined with the intrinsic spin-spin couplings, this enables universal control over the entire spin system.

\section{State Preparation}
The natural state of the experimental system differs from the initial state required for our experiment, so a dedicated preparation procedure is necessary. Specifically, we need a mixed coherent state for the system and a thermal state for the bath to initiate the collision model. This section details how the desired state is prepared. The overall procedure is divided into two parts: first, initializing a pseudo-pure state (PPS) by spatial averaging; second, preparing the mixed coherent state by temporal averaging.

We first generate the PPS from the system’s natural thermal state. A PPS is a special mixed state that is typically treated as a pure state in NMR experiments. For our four-spin molecular system, the PPS is defined as  
\begin{equation}
\renewcommand{\theequation}{S\arabic{equation}}
    \rho_{PPS}=\frac{1-\epsilon}{16}I+\epsilon\left|0000\right>\left<0000\right|,
\end{equation}
where $I$ is the $16 \times 16$ identity matrix, and the polarization $\epsilon \sim 10^{-5}$ under our experimental conditions. The identity component does not affect observable signals or the dynamics under standard NMR operations, including rf pulses and free Hamiltonian evolution. Therefore, the PPS effectively behaves like the pure state $\left|0000\right>\left<0000\right|$ up to a scaling factor. We use the spatial averaging method to obtain this state from the system’s thermal state.

\begin{figure}[htbp]
    \centering
    \includegraphics[width=1\linewidth]{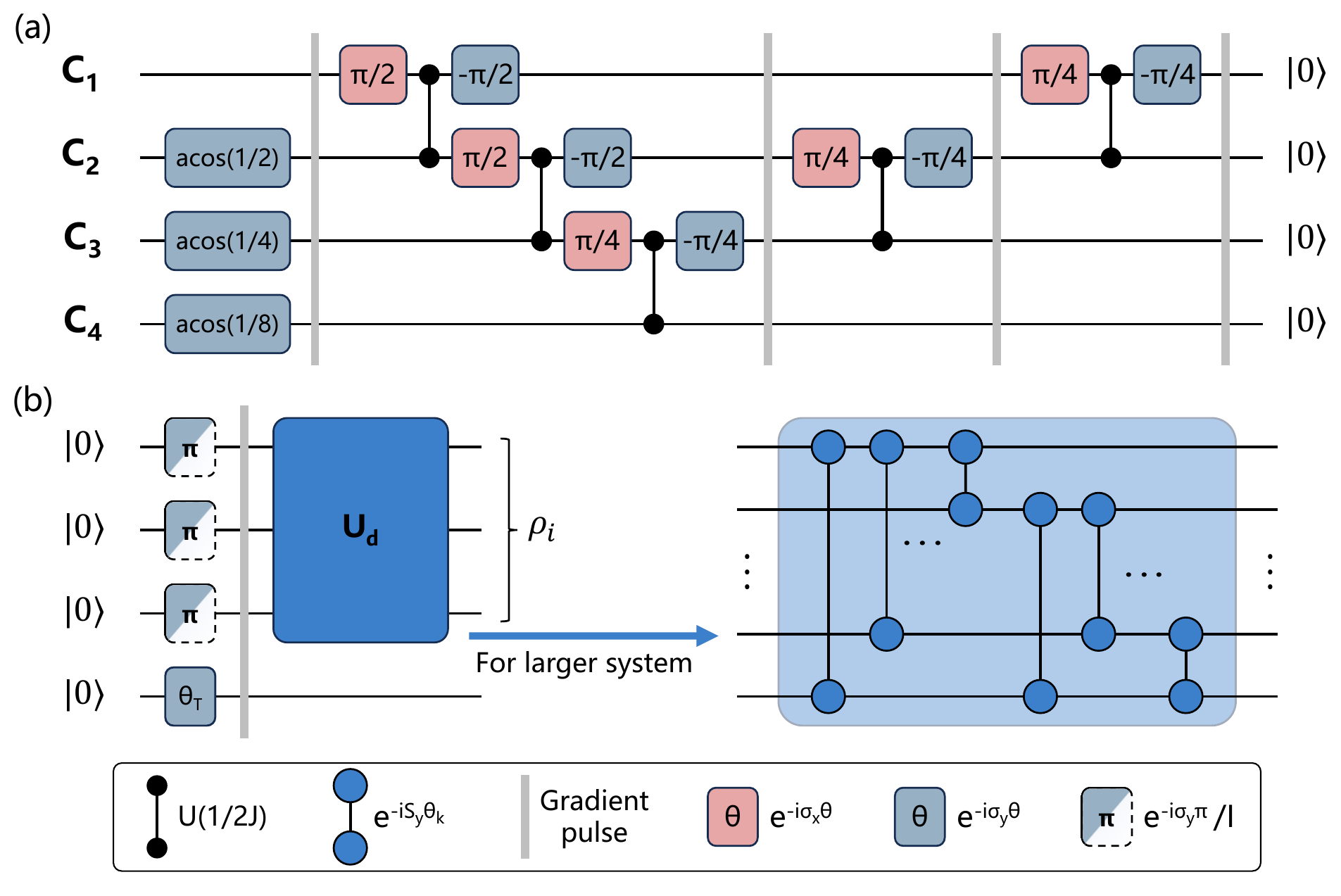}
    \caption{Circuit to prepare initial thermal state.}
    \label{circuit}
\end{figure}

Spatial averaging is a standard approach to prepare a PPS by applying magnetic field gradients to selectively suppress unwanted off-diagonal elements in the density matrix. All experiments are performed at room temperature. The thermal equilibrium state of the four-spin system is given by
\begin{equation}
\renewcommand{\theequation}{S\arabic{equation}}
    \rho_{eq}=\frac{I}{2^4}+\epsilon\sum^4_{i=k}\sigma_z^k.
\end{equation}
We use the pulse sequence shown in Figure~\ref{circuit}(a). Ignoring the identity term, the changes of the density operator during this sequence is 
\begin{equation}
\renewcommand{\theequation}{S\arabic{equation}}
    \begin{aligned}    \sigma_z^1+\sigma_z^2+\sigma_z^3+\sigma_z^4&\xrightarrow{U_{p1},\mathrm{G}z}\sigma_z^1+\frac{1}{2}\sigma_z^2+\frac{1}{4}\sigma_z^3+\frac{1}{8}\sigma_z^4\\
    &\xrightarrow{U_{p2},\mathrm{G}z}\frac{1}{2}(\sigma_z^1\sigma_z^2\sigma_z^3+\sigma_z^1\sigma_z^2\sigma_z^3\sigma_z^4)+\frac{1}{4}(\sigma_z^2\sigma_z^3+\sigma_z^2\sigma_z^3\sigma_z^4)+\frac{1}{8}(\sigma_z^3\sigma_z^4+\sigma_z^4)\\
    &\xrightarrow{U_{p3},\mathrm{G}z}\frac{1}{4}(\sigma_z^1\sigma_z^2\sigma_z^3+\sigma_z^1\sigma_z^2+\sigma_z^1\sigma_z^2\sigma_z^3\sigma_z^4+\sigma_z^1\sigma_z^2\sigma_z^4)\\&\qquad\qquad+\frac{1}{8}(\sigma_z^2+\sigma_z^2\sigma_z^3+\sigma_z^2\sigma_z^3\sigma_z^4+\sigma_z^2\sigma_z^4+\sigma_z^3+\sigma_z^3\sigma_z^4+\sigma_z^4)\\
    &\xrightarrow{U_{p4},\mathrm{G}z}\frac{1}{8}(\sigma_z^1\sigma_z^2\sigma_z^3+\sigma_z^1\sigma_z^3+\sigma_z^1\sigma_z^2+\sigma_z^1+\sigma_z^1\sigma_z^2\sigma_z^3\sigma_z^4+\sigma_z^1\sigma_z^3\sigma_z^4+\sigma_z^1\sigma_z^2\sigma_z^4\\&\qquad\qquad+\sigma_z^1\sigma_z^4+\sigma_z^2+\sigma_z^2\sigma_z^3+\sigma_z^2\sigma_z^3\sigma_z^4+\sigma_z^2\sigma_z^4+\sigma_z^3+\sigma_z^3\sigma_z^4+\sigma_z^4).
    \end{aligned}
\end{equation}
The final result is equivalent to $\bigotimes_{k=1}^4 (I_2 + \sigma^k_z)/8 = \left|0000\right>\left<0000\right|/8$, with $I_2$ the $2 \times 2$ identity matrix, yielding the desired PPS.

Next, we prepare the mixed coherent state from this PPS. Any mixed coherent state can be diagonalized via a unitary transformation, expressed as $\rho_i = U_d \rho_d U_d^\dagger$. Practically, this is implemented by independently controlling the population distribution (the diagonal elements) and then applying a unitary $U_d$ to add coherence. The diagonal state $\rho_d$ is obtained by averaging over multiple experiments prepared in different population bases. The coherence is introduced by applying an optimized rf pulse sequence that implements $U_d$. We use an optimal-control algorithm to design these pulses, achieving a simulation fidelity exceeding 99.5\% within a total pulse length of 50~ms. Figure~\ref{circuit}(b) shows this preparation process. This method can be extended to larger systems by replacing the global unitary with appropriate two-qubit gates.

\section{Measurement}
In an NMR quantum processor, the experimental sample consists not of a single molecule, but rather of an ensemble of a large number of identical molecules. Therefore, all measurements in the NMR system yield ensemble-averaged results, from which the spin state of an individual molecule can be inferred.

\begin{figure}[htbp]
    \centering
    \includegraphics[width=1\linewidth]{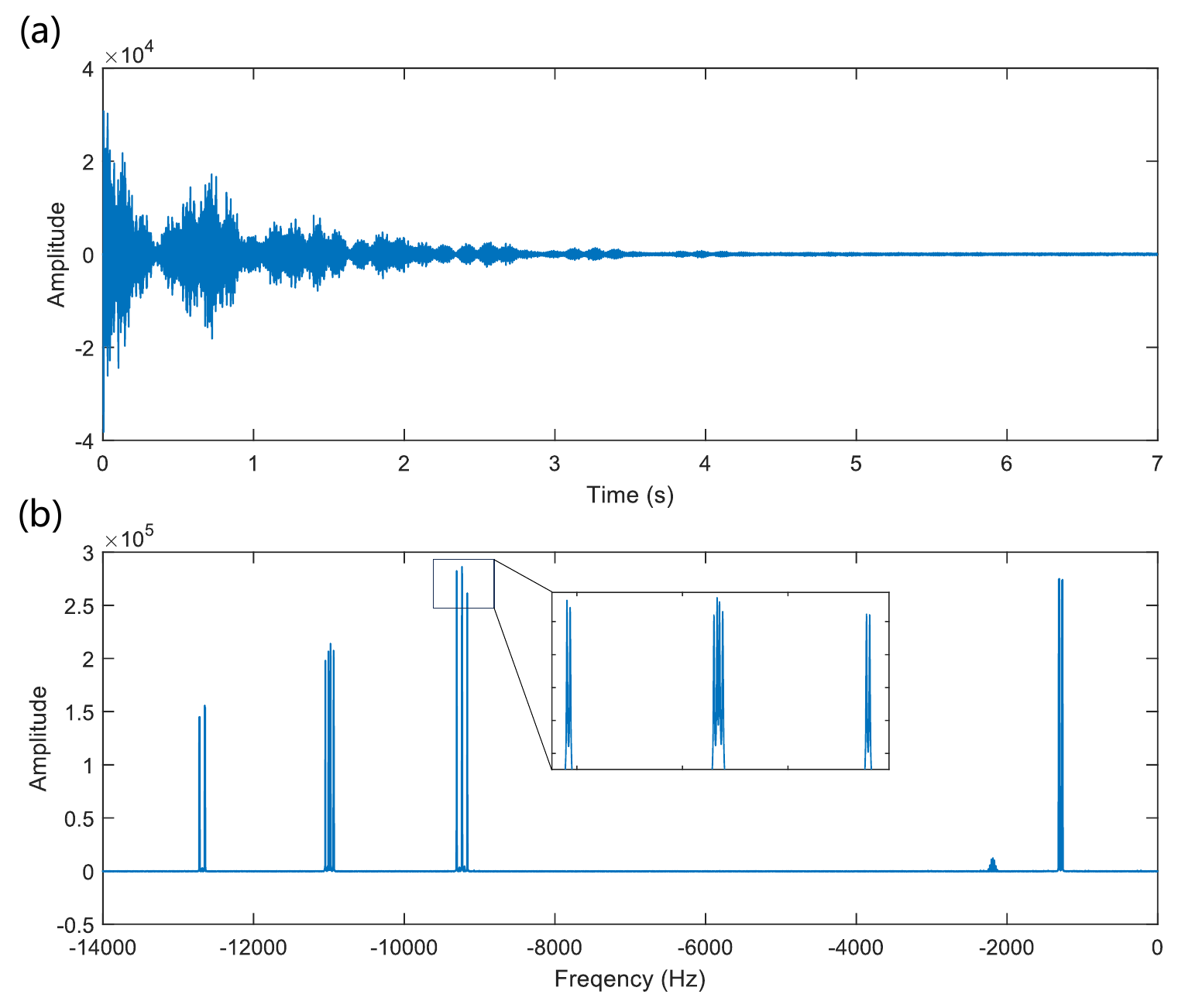}
    \caption{Results of detection in NMR system.}
    \label{thermal}
\end{figure}

During measurement, the nuclear spins precess around the static magnetic field $B_0$ along the $z$-axis and gradually relax back to thermal equilibrium. This precession induces an oscillating electrical signal in the transverse ($x$ and $y$) directions, which is detected by radio-frequency (rf) detection coils surrounding the sample. These coils capture the signal in the time domain, known as the free induction decay (FID). The measured FID signal can be written as 
\begin{equation}
\renewcommand{\theequation}{S\arabic{equation}}
    A(t)=\mathrm{Tr}[e^{-i\mathcal{H}_{\text{NMR}}t}\rho e^{i\mathcal{H}_{\text{NMR}}t}\sum_k(\sigma^k_x-i\sigma^k_y)e^{-\lambda_k t}],
\end{equation}
where $\rho$ is the density matrix of the spin system being measured, and $\lambda_k = 1/T_2$ represents the natural decoherence due to spin relaxation. Figure~\ref{thermal}(a) shows a typical FID signal for the thermal equilibrium state of the crotonic acid molecule.

However, the information contained in the FID signal is not directly apparent and requires transformation to the frequency domain for clear interpretation. This is done via a Fourier transform, which converts the time-domain signal into a frequency-domain NMR spectrum where the system information becomes more accessible. For the crotonic acid molecule used as a four-qubit NMR quantum processor, there are 32 distinct spectral lines in its spectrum, and the signal of each spin typically contains eight peaks at different frequencies due to nuclear spin couplings (Figure~\ref{thermal}(b)). The coupling strength can be determined from the frequency spacing between these peaks. According to the FID expression, the signal in the time domain includes both real and imaginary components, which correspond to the signals along the $x$ and $y$ axes and encode the expectation values of the Pauli matrices $\sigma_x$ and $\sigma_y$ for each observed spin, respectively. Thus, NMR allows measurement of the expectation values of single-quantum coherence operators composed of $\sigma_x$ or $\sigma_y$ for the target qubit and $\sigma_z$ or $I$ for the other qubits.

To measure the internal energy of each subsystem under its local Hamiltonian $H_k = \frac{\delta}{2} \sigma^k_z$ in our experiment, we focus on the longitudinal magnetization observables $\sigma_z^1III$, $I\sigma_z^2II$, and $II\sigma_z^3I$ (where the last spin represents the bath particle). The energy is then obtained as
\begin{equation}
\renewcommand{\theequation}{S\arabic{equation}}
    E_k=\mathrm{Tr}(\rho H_k)=\frac{\delta}{2}\mathrm{Tr}(\rho \sigma_z^k).
\end{equation}
However, the NMR spectrometer directly detects only transverse ($\sigma_x$ or $\sigma_y$) components. To access the longitudinal ($\sigma_z$) information, we apply a rotation pulse that maps the $z$-component onto the $x$-axis. Specifically, a rotation $R^k_y(\pi/2) = e^{-i \sigma^k_y \pi/2}$ transforms the $\sigma_z^k$ observable into $\sigma_x^k$, yielding
\begin{equation}
\renewcommand{\theequation}{S\arabic{equation}}
    \mathrm{Tr}(\rho_k \sigma^k_z)=\mathrm{Tr}\{R^k_y(\pi/2)\rho [R_y^k(\pi/2)]^{-1} \sigma_x^k\}.
\end{equation}

Furthermore, determining the system’s free energy requires knowledge of the full density matrix, not just the individual $\sigma_z$ expectations. We perform full quantum state tomography on the four-qubit system to reconstruct all relevant information. Complete tomography involves extracting 255 independent density matrix elements. To ensure robustness, we design 17 measurement settings, each involving carefully chosen $\pi$ rotations along specific Pauli operator combinations:  
$IIII$, $\sigma^1_x \sigma^2_x \sigma^3_x \sigma^4_x$, $II \sigma^3_y \sigma^4_y$, $\sigma^1_y \sigma^2_y \sigma^3_x \sigma^4_x$, $III \sigma^4_y$, $\sigma^1_x \sigma^2_y \sigma^3_x \sigma^4_x$, $\sigma^1_y \sigma^2_x \sigma^3_y I$, $I \sigma^2_x \sigma^3_y I$, $III \sigma^4_x$, $\sigma^1_x I \sigma^3_y \sigma^4_y$, $\sigma^1_y \sigma^2_x II$, $\sigma^1_y \sigma^2_y \sigma^3_x \sigma^4_y$, $\sigma^1_x \sigma^2_y \sigma^3_x I$, $II \sigma^3_y \sigma^4_x$, $I \sigma^2_x I \sigma^4_y$, $II \sigma^3_x I$, and $I \sigma^2_y I \sigma^4_y$. By combining the results of these 17 experiments, we fully reconstruct the system’s density matrix. After tracing the last spin, we can obtain the complete state of the system. Then, the free energy is computed via
\begin{equation}
\renewcommand{\theequation}{S\arabic{equation}}
    F(\rho) = \mathrm{Tr} (\rho H) + T\, \mathrm{Tr} (\rho \ln \rho),
\end{equation}
where $T$ is the bath temperature.

\section{Emulating the action of an infinite-particle reservoir}
In the ideal theoretical framework, the thermal reservoir is modeled as an infinite sequence of identical thermal particles. This assumption ensures that the reservoir remains effectively unchanged during the interaction process, thereby justifying its role as a stable heat bath in thermal contact. However, in the experimental setting, the available resources are inevitably limited. The number of controllable qubits (or particles) is finite. 

\begin{figure}[htbp]
    \centering
    \includegraphics[width=0.9\linewidth]{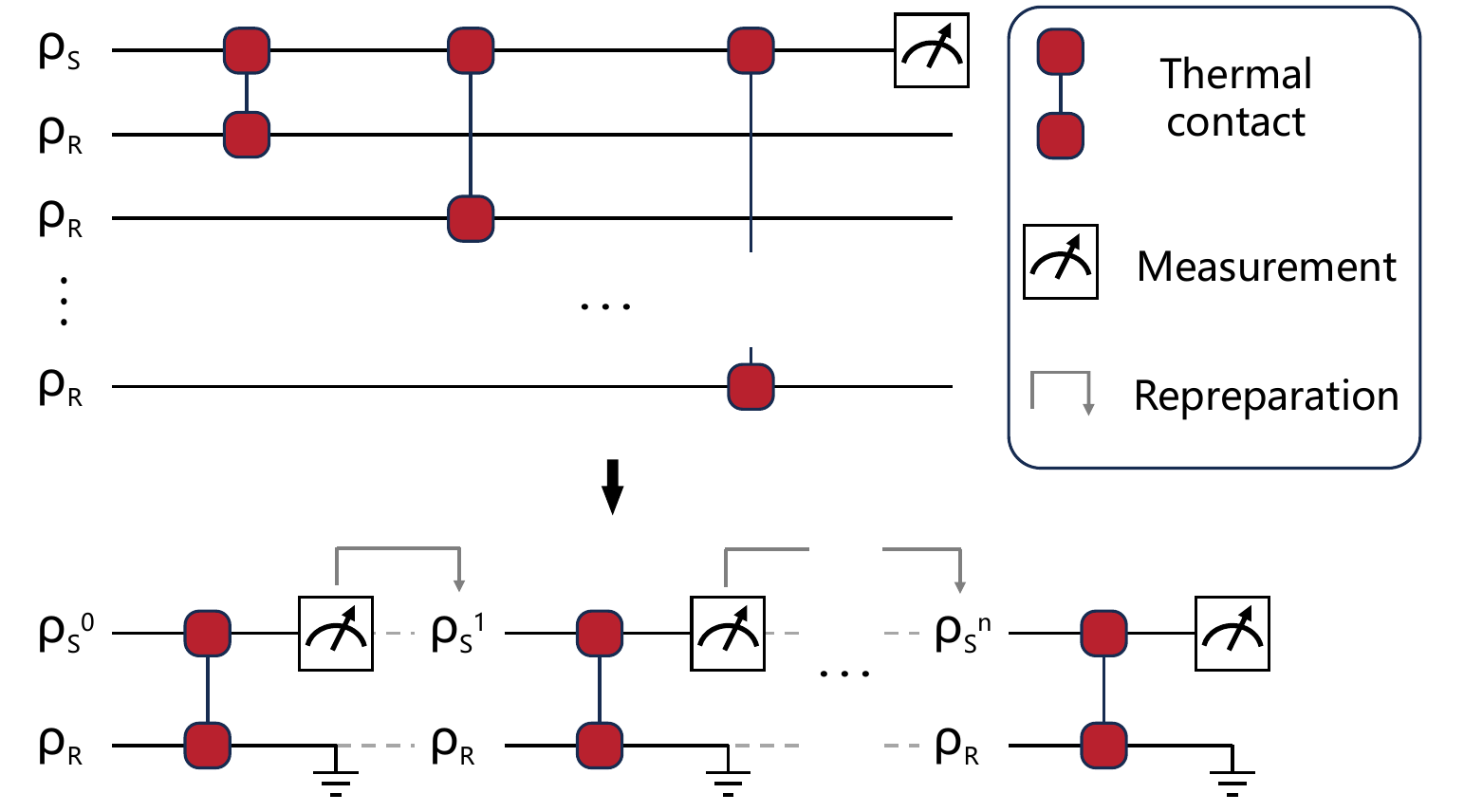}
    \caption{Experimental implementation of emulating interactions with an infinite number of heat particles by preparing a new particle state, re-preparing the system state, and repeatedly interacting with a single particle.}\label{RCircuit}
\end{figure}

To address this limitation, we emulate the action of a reservoir with infinite particles through a repeated procedure. In each run of the experiment, we implement the interaction between the system and a single thermal particle, followed by measurement of the system’s state. In the subsequent run, we reprepare the system state and initialize a fresh thermal particle to realize the next interaction. By repeating this process multiple times, each collision effectively involves a newly thermalized particle, so that the reservoir is continuously renewed. This scheme is illustrated in Fig.~\ref{RCircuit}. In this way, the sequence of repeated interactions faithfully simulates the effect of an infinite thermal bath.

\section{Work consumption in experiment}
Here we show that no net work is consumed during the entire thermal contact process, ensuring that no additional resources are introduced in our experiment.

For the whole thermal contact process, we assume the system interacts with a heat bath containing infinite particles. However, due to the limitation experiment system, a heat bath with infinite particles is unrealizable. To address this limitation, we emulate the action of a reservoir with infinite particles through a repeated procedure as explained in Fig.~\ref{RCircuit}. In this simulation, the energy change during particle replacement can be neglected. Thus, both work and heat can be regarded as zero. Moreover, in the theoretical framework, the replacement involves no work or heat exchange, consistent with the collision model. Therefore, there is no need to account for work or heat in this replacement process, which agrees with our experimental analysis.

\begin{figure}[htbp]
    \centering
    \includegraphics[width=1\linewidth]{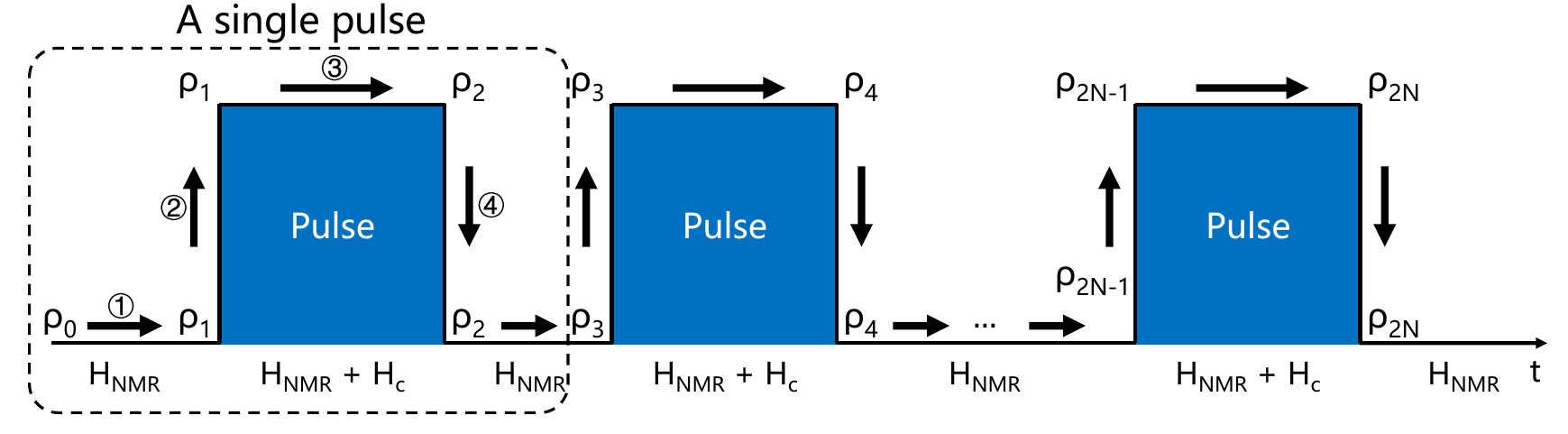}
    \caption{Sequence of pulses to achieve interaction. A single pulse with four phases from \textcircled{1} to \textcircled{4} is marked using a dotted line.}\label{PulseSequence}
\end{figure}

As for the repeated interaction procedure, also no external work is injected into the system. In our experiments, the interaction is achieved by applying a sequence of pulses, as depicted in Fig.~\ref{PulseSequence}. Each application of a pulse consists of four distinct phases, characterized by varying parameters such as the duration of intervals and pulse parameters, which are tuned to guide the system's evolution towards the desired operation. These phases include two intervals of free evolution and two intervals where the Hamiltonian is altered, sequentially labeled from \textcircled{1} to \textcircled{4}.

During the free evolution phase \textcircled{1}, no pulse is applied, meaning there is no external control field. As a result, the system evolves according to the intrinsic Hamiltonian of the molecule, $\mathcal{H}_{\text{NMR}}$. Since $\mathcal{H}_{\text{NMR}}$ is time-independent, no work is done during this phase, as defined by $dW=\mathrm{Tr}(\rho d\mathcal{H})$. The heat exchange during this phase is given by $dQ=\mathrm{Tr}(\mathcal{H} d\rho)$. The heat transfer in phase \textcircled{1} can be expressed as

\begin{equation}
\begin{aligned}
    dQ_1&=\mathrm{Tr}\left[ \mathcal{H}_\mathrm{NMR}(\rho_1-\rho_0) \right]\\[5pt]
    &=\mathrm{Tr}(\mathcal{H}_\mathrm{NMR}e^{-i\mathcal{H}_\mathrm{NMR}t_1}\rho_0 e^{i\mathcal{H}_\mathrm{NMR}t_1})-\mathrm{Tr}(\mathcal{H}_\mathrm{NMR}\rho_0).
\end{aligned}
\end{equation}

Here, $\rho_0$ is the initial state, $t_1$ is the free evolution time, and $\rho_1 = e^{-i\mathcal{H}_\mathrm{NMR}t_1}\rho_0 e^{i\mathcal{H}_\mathrm{NMR}t_1}$ is the state at the end of the phase. The result of this calculation is zero, i.e., $dQ_1=0$, consistent with the commutation relation $[\mathcal{H}, e^{-i\mathcal{H}t}] = 0$.

In phase \textcircled{2}, the Hamiltonian is altered rapidly, but the state’s variation is negligible, preventing any heat exchange. The introduction of an external control field $\mathcal{H}_c$ performs work on the system, as shown by

\begin{equation}
    dW_1=\mathrm{Tr}\left\{ \left[(\mathcal{H}_\mathrm{NMR}+\mathcal{H}_c)-\mathcal{H}_\mathrm{NMR}\right]\rho_1\right\}=\mathrm{Tr}(\mathcal{H}_c\rho_1).
\end{equation}

The third phase \textcircled{3} is similar to the first, involving free evolution, but now with the Hamiltonian including the control field $\mathcal{H}_c$. Since this Hamiltonian is also time-independent, no heat transfer occurs, as shown by

\begin{equation}
dQ_2 = \mathrm{Tr}\left[ (\mathcal{H}_\mathrm{NMR} + \mathcal{H}_c)(\rho_2 - \rho_1) \right] = 0.
\end{equation}

In the final phase \textcircled{4}, the control field is turned off, stabilizing the state $\rho_2$. Like the second phase, this also leads to no heat exchange. The work done on the system is given by

\begin{equation}
    dW_2=\mathrm{Tr}\left\{ \left[\mathcal{H}_\mathrm{NMR} -(\mathcal{H}_\mathrm{NMR} +\mathcal{H}_c)\right]\rho_2\right\}= -\mathrm{Tr}(\mathcal{H}_c\rho_2).
\end{equation}

Considering all four phases, the net change in internal energy for a single pulse is

\begin{equation}
dE_1 = dQ_1 + dW_1 + dQ_2 + dW_2 = \mathrm{Tr}\left[\mathcal{H}_c (\rho_1 - \rho_2)\right].
\end{equation}

This reflects the total work done on the system throughout the four phases. In the ideal case (neglecting experimental errors), there is no heat exchange during the pulse application. Therefore, the net change in internal energy equals the total work consumption. Summing the energy changes across the entire pulse sequence yields

\begin{equation}
\Delta E = \Delta W + \Delta Q = \Delta W.
\end{equation}

The interaction satisfies the first law of energy conservation in our framework, which means that $\Delta E=0$ holds. Therefore, no external work is injected for realizing the interaction. Moreover, this also ensures that no additional work is introduced during the entire thermal contact process.
\color{black}

\section{Resource analysis}
Let us consider the example in the main text, in which the compound system $S$ is tripartite and consists of $S_1$, $S_2$ and  $S_3$. The bath and the compound system are initially in a product state. $S_1$, $S_2$ and  $S_3$ are each locally in the same thermal state, but there is coherence between them—meaning that the compound system is not thermal and thus serves as the resource in our scenario. The dynamics of the process are characterized by a collision model: the heat bath $R$ collides with $S_1$, then with $S_2$, and finally with $S_3$.

Based on collisions, this process can be divided into three steps: (1) the first collision between $S_1$ and $R$. The correlations between $S_2$ and $R$ are created in this collision.  (2) the second collision between $S_2$ and $R$. This step consumes the correlations between $S_2$ and $R$ to reverse the heat flow. (3) the third collision between $S_3$ and $R$. This step consumes the correlations between $S_3$ and $R$ to reverse the heat flow.  The resource analysis is given below. Note that, in the following, $'$, $''$and $'''$ represent the quantities after the first, second and third collisions, respectively.

(1)The first collision. We present a relation between the free energy consumed in $S$ after the first collision and the mutual information created between $S_2$ and $R$. The free energy change of $S$ is given by

\begin{equation}\label{A1}
\renewcommand{\theequation}{S\arabic{equation}}
    \begin{aligned} 
\Delta F'(\rho_{S})&=\mathrm{Tr}[H_{S} (\rho'_{S} -\rho_{S} )] - kT_R [S(\rho'_{S}) -S(\rho_{S})]\\
&=\mathrm{Tr} [H_{S} (\rho'_{S} -\rho_{S} )] - kT_R\{ I'(S: R) + S(\rho'_{R}|| \rho_R ) + \frac{1}{kT_R} \mathrm{Tr} [H_R(\rho_R - \rho'_R) ]\}\\
&=-kT_R I'(S: R) -kT_R S(\rho'_{R}|| \rho_R )\\
&\leq -kT_R I'(S:R) \\
&\leq -kT_R I'(S_2:R),
    \end{aligned}
\end{equation}
where $H_{S}=\sum_{k=1}^{3}H_{k}$ and the third equality is due to energy conservation and the last inequality results from the strong subadditivity of quantum entropy. We therefore proved that the mutual information between $S_2$ and $R$ after the first collision is upper bounded by $\beta_R\Delta \widetilde{F}'(\rho_{S})=-\beta_R\Delta F'(\rho_{S} )$.

(2)The second collision. To account for the reversal of the heat direction, we then need to consider systems $S_2$ and $R$ in the second collision. After the second collision, the consumption of mutual information between $S_2$ and $R$ can be calculated as
\begin{equation}\label{dI}
    \begin{aligned}
\Delta \widetilde{I}''(S_2:R)&=I'(S_2:R)-I''(S_2:R)=-\Delta S(\rho_{2})-\Delta S(\rho_{R})+\Delta S(\rho_{S_2R})\\
&=-\beta_{2}\mathrm{Tr}[H_{2}(\rho''_{2}-\rho'_{2})]+S(\rho''_{2}||\rho'_{2})-\beta_{R} Tr [H_{R}(\rho_R''-\rho_R')]+S(\rho_R''||\rho_R')\\
&=\left(\beta_{2}-\beta_{R}\right)\Delta E_{R,2}+S(\rho''_{2}||\rho'_{2})+S(\rho''_R||\rho'_R)\\
&\geq\left(\beta_{2}-\beta_{R}\right)\Delta E_{R,2},
    \end{aligned}
\end{equation}
where $\Delta E_{R,2}\equiv \mathrm{Tr} [H_{R}(\rho_R''-\rho_R')]$ is the energy change of R for the second interaction that also represents heat transfer from $S_2$ to $R$. Assume that $T_{2} \geq T_R$, the negative value of $\Delta E_{R,2}$ means the heat transfer from $R$ to $S_2$, i.e., the heat transferred from cold to hot. The amount of anomalous heat transfer is bounded by the decrease in mutual information between $S_2$ and $R$.

Combined with equation Eq. (\ref{A1}), we can write for the second collision
\begin{equation}\label{FF2}
\beta_R\Delta \widetilde{F}'(\rho_{S})\geq \Delta \widetilde{I}''(S_2:R)\geq\left(\beta_{2}-\beta_{R}\right)\Delta E_{R,2}.
\end{equation}
Then, a reversal of finite heat exchange between $S_2$ and $R$ implies $\left(\beta_{2}-\beta_{R}\right)\Delta E_{R,2}>0$, which means a strict consumption of the free energy of $S$. 

Similar to Eq. (\ref{A1}), the free energy change of $S$ after the first two collisions can be written as
\begin{equation}\label{FF4}
    \begin{aligned}
\Delta F''(\rho_{S})&=\mathrm{Tr} [H_{S} (\rho''_{S} -\rho_{S} )] - kT_R [S(\rho''_{S}) -S(\rho_{S})]\\
&= \mathrm{Tr} [H_{S} (\rho''_{S} -\rho_{S} )] - kT_R\{I''(S: R) + S(\rho''_{R}|| \rho_R ) + \frac{1}{kT_R} \mathrm{Tr} H_R[(\rho_R - \rho''_R) ]\} \\
&= -kT_R  I''(S: R) -kT_R S(\rho''_{R}|| \rho_R ) \\
&\leq -kT_R I''(S:R) \\
&\leq -kT_R I''(S_3:R),
    \end{aligned}
\end{equation}
showing the mutual information between $S_3$ and $R$  after the second collision is upper bounded by $\beta_R\Delta \widetilde{F}''(\rho_{S})=-\beta_R\Delta F''(\rho_{S} )$.

(3)The third collision. After the collision $S_3-R$, we can write
\begin{equation}\label{dI2}
    \begin{aligned}
\Delta \widetilde{I}'''(S_3:R)&=I''(S_3:R)-I'''(S_3:R)\\
&=\left(\beta_{3}-\beta_{R}\right)\Delta E_{R,3}+S(\rho'''_{3}||\rho''_{3})+S(\rho_R'''||\rho_R'')\\
&\geq\left(\beta_{3}-\beta_{R}\right)\Delta E_{R,3},
    \end{aligned}
\end{equation}
and
\begin{equation}\label{FF2}
\beta_R\Delta \widetilde{F}''(\rho_{S})\geq \Delta \widetilde{I}'''(S_3:R)\geq\left(\beta_{3}-\beta_{R}\right)\Delta E_{R,3},
\end{equation}
where $\Delta E_{R,3}\equiv \mathrm{Tr} [H_{R}(\rho_R'''-\rho_R'')]$ is the energy change of R for the third interaction that also represents heat transfer from $S_3$ to $R$. Then, a reversal of finite heat exchange between $S_3$ and $R$ implies $\left(\beta_{3}-\beta_{R}\right)\Delta E_{R,3}>0$, which means a strict consumption of the free energy of $S$.

Moreover, from the global perspective by considering all collision as a unitary evolution, we can understand the function of coherence more directly. We employ the relative entropy of coherence to quantify coherence, which allows us to directly capture changes during the heat flow. The correlation between $S$ and $R$ is quantified by their mutual information:

\begin{equation}
\begin{aligned}
   \Delta I(S:R) &= \Delta S(\rho_{S}) + \Delta S(\rho_{R}) - \Delta S(\rho_{SR}) \\[5pt]
    &= \Delta S[\mathcal{D}_\mathrm{H}(\rho_S)] - \Delta C(\rho_S) + \Delta S(\rho_{R}) \\[5pt]
    &= \beta_S \mathrm{Tr} \left\{ H_{S} \left[\mathcal{D}_\mathrm{H}(\rho'_S) - \mathcal{D}_\mathrm{H}(\rho_S)\right] \right\} - S(\mathcal{D}_\mathrm{H} (\rho'_S) || \mathcal{D}_\mathrm{H}(\rho_S)) - \Delta C(\rho_S) \\[5pt]
    & \quad + \beta_R \mathrm{Tr} \left[ H_{R} (\rho_R' - \rho_R) \right] - S(\rho_R' || \rho_R) \\[5pt]
    & = -\Delta C(\rho_S) + (\beta_R - \beta_S) \Delta E_R - S(\mathcal{D}_\mathrm{H} (\rho'_S) || \mathcal{D}_\mathrm{H}(\rho_S)) - S(\rho_R' || \rho_R).
\end{aligned}
\end{equation}

For the second equals sign, the coherent part of the system is separated and measured by the relative entropy of coherence, defined as \( C(\rho) = S[\mathcal{D}_\mathrm{H}(\rho_S)] - S(\rho_S) \), where \( S(\rho_S) \) is the von Neumann entropy and \( \mathcal{D}_\mathrm{H}(\rho_S) \) denotes the state obtained from \( \rho_S \) by removing all off-diagonal elements in the basis that diagonalizes the Hamiltonian. The coherent term does not contribute energy since \( \mathrm{Tr}[H_S \mathcal{D}_\mathrm{H}(\rho_S)] = \mathrm{Tr}(H_S \rho_S) \). The term \( \Delta S(\rho_{SR}) = 0 \) is removed because unitary evolution does not change entropy. 

For the third equals sign, the entropy changes of $S$ and $R$ are replaced using the equation 
\begin{equation}
 S(\rho' || \rho) = S(\rho' || \rho) - S(\rho || \rho) = -S(\rho') + S(\rho) + \beta \mathrm{Tr}[H(\rho' - \rho)]. 
\end{equation}
This equation holds as $S(\rho || \rho)  \equiv 0$. Here, the initial state is assumed to be a Gibbs state \(\rho = e^{-\beta H} / Z\), where \( \beta \) is the inverse temperature and \( Z \) is the partition function. 

For the final equals sign, under the first law of thermodynamics, heat flow can be represented by the increase or decrease of the bath particle's energy as 
\begin{equation}
 \Delta E_R = \mathrm{Tr} [H_{R} (\rho_R' - \rho_R)] = -\mathrm{Tr} \left\{ H_S \left[\mathcal{D}_\mathrm{H}(\rho'_S) - \mathcal{D}_\mathrm{H}(\rho_S)\right] \right\} .
 \end{equation}

As a result, we obtain the equation to describe heat flow between $S$ and $R$, considering coherence:

\begin{equation}
    (\beta_R - \beta_S) \Delta E_R = \Delta I(S:R) + \Delta C(\rho_S) + S(\mathcal{D}_\mathrm{H} (\rho'_S) || \mathcal{D}_\mathrm{H}(\rho_S)) + S(\rho_R' || \rho_R).
\end{equation}

The last two terms on the right side are non-negative because they involve relative entropy, so we obtain the inequality:

\begin{equation}
    (\beta_R - \beta_S) \Delta E_R \ge \Delta I(S:R) + \Delta C(\rho_S).
\end{equation}

Without loss of generality, consider the case that the system's temperature is higher than that of the bath, i.e., \( \beta_R - \beta_S > 0 \). In this case, normal means the heat flows from the system to the bath, so the energy of the bath particle increases (\( \Delta E_R > 0 \)). Thus, \( (\beta_R - \beta_S) \Delta E_R > 0 \) just indicates normal heat flow. Meanwhile, \( (\beta_R - \beta_S) \Delta E_R < 0 \) indicates reversed heat flow.

\section{State changes in interaction with one bath particle}
To demonstrate our analysis, we design experiments to study the changes in resources during cascade interactions. We consider a model consisting of two subsystems, $S_1$ and $S_2$, and a heat bath particle $R$. Each subsystem has Hamiltonian $H_k = \frac{\delta}{2} \sigma_z^k$, while the bath particle has Hamiltonian $H_R = \frac{\delta}{2} \sigma_z^R$. In the eigenbasis of the system, the initial coherent state is given by
\begin{equation}    
    \rho_{12}^0 = \rho_1^\mathrm{th} \otimes \rho_2^\mathrm{th} +  \lambda e^{i\alpha}\chi + \text{h.c.},
\end{equation}
where $\rho_k^\mathrm{th} = e^{-\beta_k H_k} / \mathrm{Tr}(e^{-\beta_k H_k})$ are local Gibbs states, with $\beta_k = 1/T_k$ denoting the inverse temperature. The term $\chi$ is constructed to introduce off-diagonal coherence between the degenerate energy levels $\left|ge\right>$ and $\left|eg\right>$, where $\left|g\right>$ and $\left|e\right>$ represent the ground and excited states of the subsystem, respectively. The parameters $\lambda$ and $\alpha$ characterize the strength and phase of the coherence.

For the collision model with cascade interaction, each subsystem interacts sequentially with a single heat bath particle, forming the system–bath interaction. The local interaction is governed by the Hamiltonian 
\begin{equation}
H_{kR} = g(\sigma_+^k \sigma_-^R + \sigma_-^k \sigma_+^R),
\end{equation} 
where \(\sigma_\pm^k\) and \(\sigma_\pm^R\) are the standard ladder operators for the subsystem and the bath particle, respectively, and \(g\) denotes the interaction strength.

\begin{figure}[htbp]
    \centering
    \includegraphics[width=0.9\linewidth]{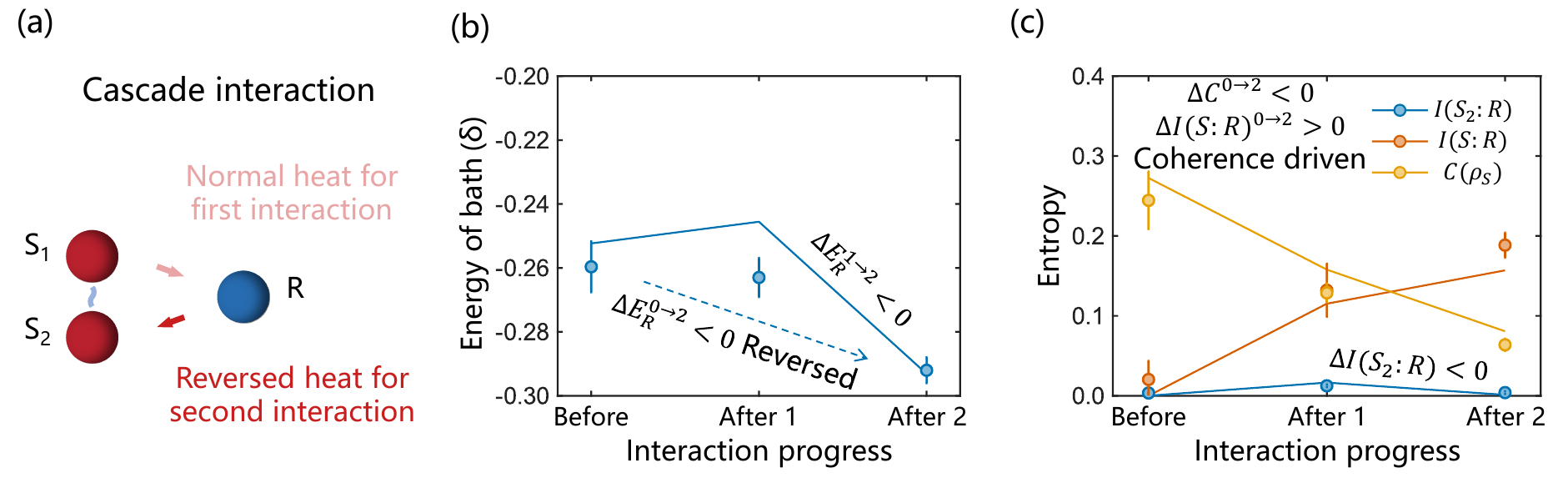}
    \caption{Changes during cascade interaction. Decrease in bath's energy indicates the reversed heat flow.}\label{cascade}
\end{figure}

We record the bath particle's energy \( E_R\), mutual information \(I(S:R)\), and relative entropy of coherence \(C(\rho_S)\) during the interaction. The parameters are chosen as in our initial manuscript: \(T_1 = T_2 = \delta/k_\mathrm{B}\), \(T_R = 0.9\delta/k_\mathrm{B}\), and \(g = 20\delta\). To enhance the observable changes, we increase the action time and coherence strength (\(\tau = 0.03\delta^{-1}\) and \(\lambda = 1\) with \(\alpha = \pi\)). For the case of heat flow reversal, the simulated and experimental results are presented in Fig.~\ref{cascade}.

Before the interaction, the system possesses only coherence without any system–bath correlations, i.e., \(C(\rho_S) > 0\) and \(I(S:R) = 0\). Focusing on the subsystem \(S_2\) and the bath particle \(R\), their correlation, as quantified by the mutual information \(I(S_2:R)\), is generated after the first interaction. Meanwhile, the relative entropy of coherence \(C(\rho_S)\) decreases, indicating that coherence is consumed as a resource to create \(I(S_2:R)\). After the second interaction, \(C(\rho_S)\) continues to decrease, and \(I(S_2:R)\) decreases to nearly zero, while heat flow reversal occurs. These information measures are shown in Fig.~\ref{cascade}(c). At this stage, the mechanism could be interpreted using the correlation-driven picture, as you pointed out.  

However, we observe that \(C(\rho_S)\) also decreases after the second interaction, indicating that coherence continues to be consumed. Therefore, it is not convincing to claim that correlation alone fuels the reversed heat flow. To clarify this point, we consider the entire system \(S\), comprising both \(S_1\) and \(S_2\). We find that the total mutual information \(I(S:R)\) increases after both interactions; see Fig.~\ref{cascade}(c). This result clearly demonstrates that system–bath correlation is not the driving resource; rather, it is the receiver, gaining resource from coherence. Hence, it is reasonable to conclude that the heat flow reversal is primarily driven by the coherence resource.

Moreover, we demonstrate that heat flow can be reversed purely by coherence, even when no correlation is present. We consider a simultaneous interaction model in which the two subsystems, \(S_1\) and \(S_2\), interact with a single bath particle at the same time rather than sequentially. With sufficient coherence, this interaction can also induce heat flow against the temperature gradient. No correlation is established prior to the heat reversal because both interactions occur simultaneously.

\begin{figure}[htbp]
    \centering
    \includegraphics[width=0.9\linewidth]{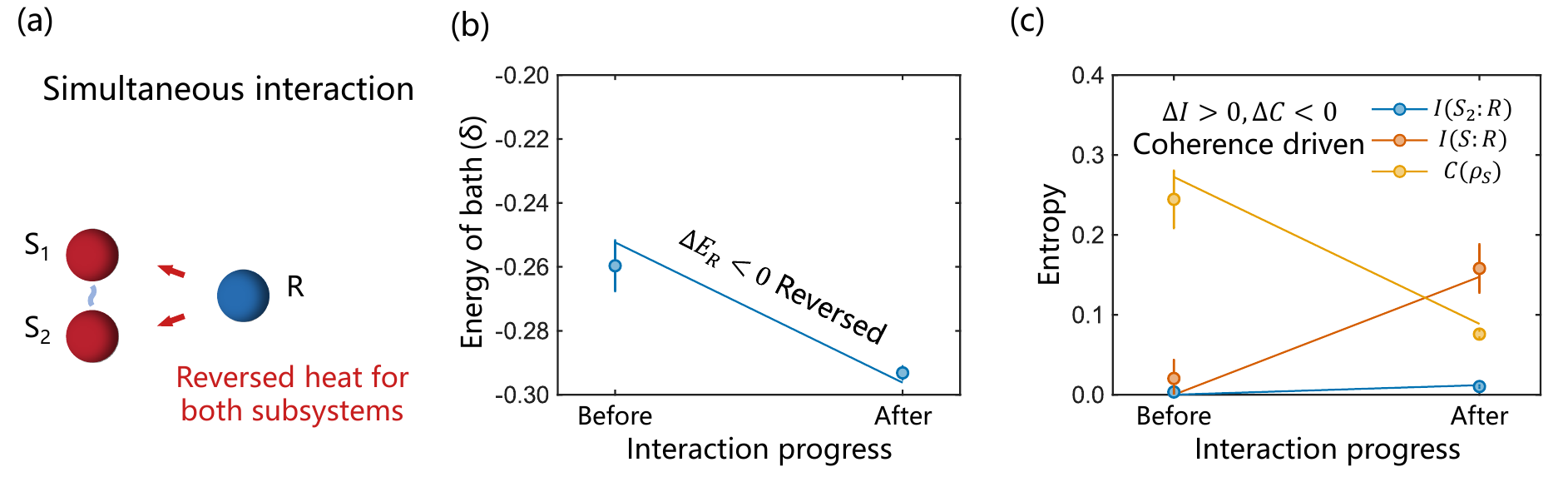}
    \caption{Changes during simultaneous interaction.}\label{simultaneous }
\end{figure}

All parameters are set identical to those used in the cascade model. We monitor the same quantities during this simultaneous interaction, and the results are presented in Fig.~\ref{simultaneous }. After the interaction, heat is transferred from the bath particle to both subsystems ($\Delta E_R < 0$, reversed heat flow shown in Fig.~\ref{simultaneous }(b)), while the mutual information also increases, i.e., \(\Delta I(S:R) > 0\) [Fig.~\ref{simultaneous }(c)]. The only consumed resource is coherence. This result clearly demonstrates that the process is driven by coherence, which cannot be accounted for by previous correlation-driven pictures.

\section{Continuous thermal contact of two-spin system}
\begin{figure}[htbp]
    \centering
    \includegraphics[width=0.9\linewidth]{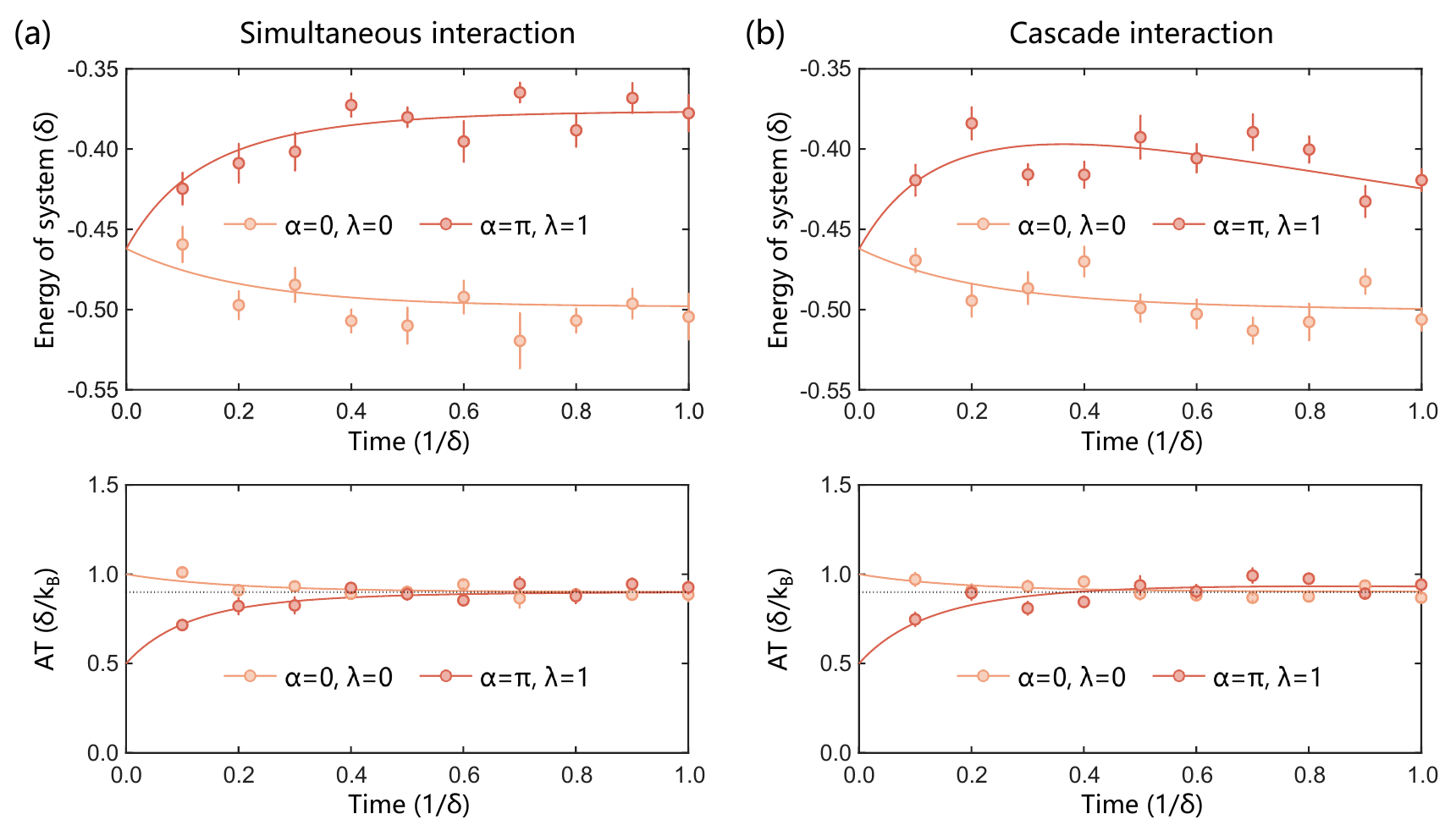}
    \caption{Energy and apparent temperature (AT) changes during thermal contact. For both the cascade and simultaneous interaction models, heat continues to transfer from the region of high AT to the region of low AT.}\label{ATnewExp}
\end{figure}

We also implement thermal contact experiments of two-spin system for both interaction types by continually resetting the heat reservoir particles, as described in our initial manuscript. The action time is set to $\tau = 0.01\delta^{-1}$ to approximate ideal thermal contact. The corresponding changes in energy and apparent temperature (AT) of the system during thermal contact are shown in Fig.~\ref{ATnewExp}. 

Heat flow is reversed in a similar manner in both interaction models, starting from identical initial states. Notably, our method of assessing heat flow via AT is also applicable to the simultaneous interaction model, demonstrating that it serves as an effective metric for capturing the influence of coherence on energy exchange. Heat continues to transfer from regions of high AT to regions of low AT.
\color{black}

\section{Apparent temperature}
Here, we derive the apparent temperature of the subsystem in the collision model with cascade interaction. After a complete interaction process from $S_{1}-R$ to $S_{n}-R$, the system and the bath particle evolve under the sequential application of $n$ evolution operators: $U_1$, $U_2$, $\cdots$ and $U_n$. The system state $\rho_S$ is then transformed into a new state $\rho_S^{n\prime}$, which is described by a completely positive and trace-preserving (CPTP) map $\Phi$ acting on $\rho_S$:
\begin{equation}
    \rho_S^{n\prime}=\Phi(\rho_S)=\mathrm{Tr}_R(U_n \cdot\cdot\cdot U_2 U_1 \rho_{SR} U_1^\dagger U_2^\dagger \cdot\cdot\cdot U_n^\dagger),
\end{equation}
where $\rho_{SR}=\rho_S\otimes\rho_R$ is the initial joint state of the system and the heat bath particle. The bath particle is in a thermal state $\rho_R=e^{-\beta_RH_R}/\mathrm{Tr}(e^{-\beta_RH_R})$, with inverse temperature of bath $\beta_R=1/k_BT_R$.

We expand the unitary $U_k$ in series up to second order in the short interaction time $\tau$ to find
\begin{equation}
U_k\simeq \hat{\mathbb{I}}-i(H_{k} + H_R + V_{kR})\tau-\frac{1}{2}(H_{k} + H_R + V_{kR})^{2}\tau^{2}.
\end{equation}
This gives the reduced state of the system
\begin{equation}
    \rho_S^{n\prime}=\rho_S-i\tau[H_S,\rho_S]+\tau\mathcal{L}\left(\rho_{S}\right)
\end{equation}
with $H_S=\sum_{k=1}^{n}H_{k}$. Dividing both sides by $\tau$ and taking the limit of $\tau\rightarrow0$, we can derive the quantum master equation
\begin{equation}\label{QME}
\dot{\rho}_{S}=\lim_{\tau\rightarrow0}\frac{\rho_{S}^{n\prime}-\rho_{S}}{\tau}=-i\left[H_{S},\rho_{S}\right]+\mathcal{L}\left(\rho_{S}\right),
\end{equation}
where the cascaded dissipator $\mathcal{L}\left(\rho_{S}\right)$ can be decomposed into local and bipartite nonlocal terms as
\begin{equation}
    \mathcal{L}\left(\rho_{S}\right)=\sum_{k=1}^{n} \mathcal{L}_k(\rho_S)+2\sum_{l>k} \mathcal{D}_{kl}(\rho_S).
\end{equation}

When the $V_{kR}$ is given as the main text, the explicit forms of local and nonlocal dissipation terms can be written as
\begin{equation}\label{D31}
    \begin{aligned}
\mathcal{L}_{k}(\rho_{S})&=
\frac{\gamma^{+}}{2}\left(2\sigma_+^k\rho_{S}\sigma_-^k
-\left[\sigma_-^k \sigma_+^k,\rho_{S}\right]_{+}\right)\\
&+\frac{\gamma^{-}}{2}
\left(2\sigma_-^k \rho_{S}\sigma_+^k
-\left[\sigma_+^k\sigma_-^k ,\rho_{S}\right]_{+}\right),
    \end{aligned}
\end{equation}
and
\begin{equation}\label{D32}
    \begin{aligned}
\mathcal{D}_{kl}(\rho_{S})
&=\frac{\gamma^{+}}{2}\left(\sigma_+^k \left[\rho_{S}, \sigma_-^l\right]
+\left[\sigma_+^l, \rho_{S} \right]\sigma_-^k \right)\\
&+\frac{\gamma^{-}}{2}\left(\sigma_-^k \left[ \rho_{S}, \sigma_+^l\right]
+\left[\sigma_-^l, \rho_{S} \right]\sigma_+^k\right),
    \end{aligned}
\end{equation}
in which $\gamma^{+}=g^{2}\tau\langle\sigma_+^R\sigma_-^R\rangle_{R}$,
$\gamma^{-}=g^{2}\tau\langle\sigma_-^R\sigma_+^R\rangle_{R}$, and $\langle \cdot\cdot\cdot \rangle_{R}=\mathrm{Tr}[(\cdot\cdot\cdot)\rho_{R}]$.

The energy of the subsystem $S_{k}$ defined as $E_{k}=\mathrm{Tr}(\rho_{S}H_{k})$, which is used in the main text to indicate the reversal of heat flow. This is because of the energy conservation interaction between $S_{k}-R$, and then the heat current flowing from the reservoir $R$ to the subsystem $S_{k}$ can be formulated as $\dot{Q}_{k}=\dot{E}_{k}=\mathrm{Tr}(\dot{\rho}_{S}H_{k})$.
By means of the master equation describing the system's dynamics
$\dot{\rho}_{S}$ in Eq. (\ref{QME}) and the explicit dissipators in Eqs. (\ref{D31}) and (\ref{D32}),
the heat current $\dot{Q}_{k}$ can be further expressed as
\begin{equation}\label{QSicommon}
\dot{Q}_{k}=-\delta\gamma^{-}\left( P^1_k+\mathcal{C}_{k}\right)
+\delta\gamma^{+}\left(P^0_k+\mathcal{C}_{k}\right).
\end{equation}
Here, $P^0_k=\langle\sigma_-^k \sigma_+^k\rangle_{S}$ and $P^1_k=\langle\sigma_+^k \sigma_-^k\rangle_{S}$ are ground and excited state populations of subsystem $S_k$, respectively, with $\langle\cdot\cdot\cdot\rangle_{S}=\mathrm{Tr}[(\cdot\cdot\cdot)\rho_{S}]$ denoting the expectation over the system state. $\mathcal{C}_{k}$ represents the contribution from coherence between the degenerate energy levels of $S_k$ and its preceding subsystems $S_p$ with $p<k$, which we refer to as the one-way coherence of $S_k$, and takes on the form
\begin{equation}
\mathcal{C}_{k}=\sum_{p<k}\langle\sigma_-^p \sigma_+^k\rangle_{S}+\langle\sigma_+^p \sigma_-^k\rangle_{S}.
\end{equation}

By virtue of the detailed balance condition $\gamma^{+}/\gamma^{-}=e^{-\delta/T_{R}}$, the heat current $\dot{Q}_{k}$ in
Eq. (\ref{QSicommon}) can be updated to
\begin{equation}\label{QSicommon1}
\dot{Q}_{k}=\delta\gamma^{-}\left(P^0_k
+\mathcal{C}_{k}\right)
\left(e^{-\delta/ T_{R}}-e^{-\delta/AT_{k}}\right),
\end{equation}
the sign of which is determined by the difference between the temperature $T_{R}$ and the quantity $AT_{k}$. Therefore, we refer to $AT_{k}$ as the apparent temperature of $S_{k}$ with its explicit expression being given as
\begin{equation}
    AT_{k} = \delta \left( \ln \frac{P^0_k +\mathcal{C}_k }{P_k^1 + \mathcal{C}_k } \right)^{-1}.
\end{equation}

To obtain the conditions for heat flow reversal, we decompose the apparent temperature $AT_{k}$ of $S_{k}$ into
\begin{equation}
\frac{\delta}{AT_{k}}=\frac{\delta}{T_{k}}+\ln\frac{1+\mathcal{C}_{k}/P_k^0}{1+\mathcal{C}_{k}/P_k^1}.
\end{equation}
Then, the conditions for one-way coherence of $S_{k}$ enabling heat flow reversal is
\begin{equation}
\mathcal{C}_{k}<\frac{e^{\beta_k\delta}-e^{\beta_R\delta}}{(e^{\beta_k\delta}+1)(e^{\beta_R\delta}-1)}<0
\end{equation}
in the case of ${T_{k}}>{T_{R}}$, and
\begin{equation}
\mathcal{C}_{k}>\frac{e^{\beta_k\delta}-e^{\beta_R\delta}}{(e^{\beta_k\delta}+1)(e^{\beta_R\delta}-1)}>0
\end{equation}
in the case of ${T_{k}}<{T_{R}}$.

Moreover, to explain the reversed heat flow from a global perspective, we extend our derivation to obtain the apparent temperature of the entire system $S$ in the collision model with cascade interaction. The energy of the system $S$ is defined as $E_{S}=\mathrm{Tr}(\rho_{S}H_{S})$. The heat current flowing into the system from the bath is defined as $\dot{Q}_{S}=\mathrm{Tr}(\dot{\rho}_{S}H_{S})$, which can be further expressed as
\begin{equation}\label{QSt}
\dot{Q}_{S}=\sum_{k}\dot{Q}_{k}=\delta\gamma^{-}\left(\sum_{k}P^0_k
+\mathcal{C}\right)
\left(e^{-\delta/ T_{R}}-e^{-\delta/AT}\right),
\end{equation}
where $\mathcal{C}=\sum_{l\neq k}\langle\sigma_-^l \sigma_+^k\rangle_{S}+\langle\sigma_+^l \sigma_-^k\rangle_{S}$ represents the contribution of coherence in the system, and $AT$ acts as the temperature for system $S$ with specific expression
\begin{equation}\label{ATS1}
    AT=\delta \left( \ln \frac{\sum_{k}P^0_k +\mathcal{C}}{\sum_{k}P_k^1 + \mathcal{C} } \right)^{-1}.
\end{equation}
This apparent temperature of the system represented by Eq. (\ref{ATS1}) is applicable not only to cascade interaction but also to simultaneous interaction. 

Notably, AT also explains why the reversal of heat flow is driven by the consumption of internal coherence of the system, beyond the correlation-driven mechanisms discussed in previous literature. The generality of AT lies in its ability to capture both the system’s state and the influence of coherence on system’s dynamics. In our setting, the system’s state alone cannot determine the direction of heat flow; however, with knowledge of its dynamics, one can predict the future state. By comparing the thermodynamic quantities before and after the evolution, one can thus determine the direction of heat flow. This is precisely the idea behind the concept of AT. Given the interaction in our model, the heat flow can therefore be controlled through the coherence in the system’s state, as characterized by the AT.

\color{black}

\section{Modulating energy with coherence}
Coherence in the off-diagonal part can influence the system's energy under some specific operations. The amount of energy affected in this way is related to the coherence strength $C$, regardless of its phase. Here, we illustrate this influence through three models.

The first is a simple model consisting of only one spin-$\frac{1}{2}$ system described by the Hamiltonian $\mathrm{H}_1 = \delta \sigma_z / 2$, where $\sigma_{z(x,y)}$ are the Pauli operators. $\delta$ represents the energy scale. If the state initially has a coherence magnitude $C$ ($0 \le C \le 1/2$), it can be written as
\begin{equation}
\begin{aligned}
    \rho_1^\mathrm{st}=
\begin{bmatrix}
 1/2 & C\\
 C & 1/2
\end{bmatrix}.
\end{aligned}
\end{equation}
There exists a unitary operation $U_1 = e^{-i \sigma_y \pi / 2}$ that can be applied to this state, transforming it into
\begin{equation}
    \begin{bmatrix}
    1/2 & C\\
    C & 1/2
    \end{bmatrix}\xrightarrow{U_1}
    \begin{bmatrix}
    1/2-C & 0\\
    0 & 1/2+C
    \end{bmatrix}.
\end{equation}
During this process, the energy of the system increases by $C \delta$. Meanwhile, two levels with different energies are swapped, requiring work to perform the operation. Essentially, the system's energy increases due to the external work applied. Moreover, when the system becomes larger and the two swapped levels become degenerate, the scenario becomes more interesting.

The second model consists of two spins with the same energy splitting, described by the Hamiltonian $\mathrm{H}_2 = \frac{\delta}{2} (\sigma_{z}^1 + \sigma_{z}^2)$. Coherence is added in a manner similar to that used for two subsystems in the main text: coherence elements are introduced into the off-diagonal parts without affecting the local state of each spin. The amount of this coherence must also satisfy $0 \le C \le 1/2$. The initial state of this model is given by
\begin{equation}
\begin{aligned}
    \rho_2^\mathrm{st}=
    \begin{bmatrix}
    0 & 0 & 0 & 0 \\
    0 & 1/2 & C & 0 \\
    0 & C & 1/2 & 0 \\
    0 & 0 & 0 & 0 
    \end{bmatrix}.
\end{aligned}
\end{equation}
By applying a similar unitary operation that exchanges the populations of the $\left|0_1 1_2\right\rangle$ and $\left|1_1 0_2\right\rangle$ levels, the state evolves as
\begin{equation}
    \begin{bmatrix}
    0 & 0 & 0 & 0 \\
    0 & 1/2 & C & 0 \\
    0 & C & 1/2 & 0 \\
    0 & 0 & 0 & 0 
    \end{bmatrix}\xrightarrow{U_2}
    \begin{bmatrix}
    0 & 0 & 0 & 0 \\
    0 & 1/2-C & 0 & 0 \\
    0 & 0 & 1/2+C & 0 \\
    0 & 0 & 0 & 0 
    \end{bmatrix}.
\end{equation}
In this process, the first spin transfers heat $\Delta Q = C \delta$ to the second spin. No external work is required because the levels $\left|0_1 1_2\right\rangle$ and $\left|1_1 0_2\right\rangle$ are degenerate. Through this process, energy is modulated purely via coherence between different subsystems. Additionally, coherence does not have to be shared between the two heat-exchanging spins; it can reside solely in one subsystem to modulate the energy.

The final model is a simplified cascade model based on the framework discussed in the main text. This model comprises two parts: a two-spin system with coherence identical to that in the previous model, and an additional spin acting as an interaction mediator. The total Hamiltonian is $\mathrm{H}_3 = \frac{\delta}{2} (\sigma_{z}^1 + \sigma_{z}^2 + \sigma_{z}^m)$, with all spins initialized at the same temperature. In this model, the coherent levels $\left|0_1 1_2\right\rangle$ and $\left|1_1 0_2\right\rangle$ are linked via the mediator through two paths:
\begin{equation}
    \begin{aligned}
        \left|1_1 0_2 0_m\right>\to\left|0_1 0_2 1_m\right>\to\left|0_1 1_2 0_m\right>,\\
        \left|0_1 1_2 1_m\right>\to\left|1_1 1_2 0_m\right>\to\left|1_1 0_2 1_m\right>.
    \end{aligned}
\end{equation}
To utilize the initial coherence between the first and last levels, it is first transformed by suitable preparation operations and then consumed by the final interaction. This is exactly what happens during the cascade process. Throughout this process, all involved levels span a subspace of the total Hilbert space, and operators acting on these levels do not drive them outside this subspace. Therefore, the state evolution can be analyzed within this subspace.

Taking the upper path as an example, with a coherence element $C$ ($0 \le C \le \sqrt{P^{010} P^{100}}$), the reduced density matrix can be written as
\begin{equation}
    \rho_3^\mathrm{st}=
    \begin{bmatrix}
    P^{001} & 0 &  0 \\
    0 & P^{010} &  C \\
    0 & C & P^{100} 
    \end{bmatrix}.
\end{equation}
Note that the diagonal elements satisfy $P^{001} = P^{010} = P^{100}$ due to the equal initial temperatures of the three spins. Through the cascade interaction, the evolution of this reduced density matrix is
\begin{equation}
\begin{aligned}
    &\begin{bmatrix}
    P^{001} & 0 &  0 \\
    0 & P^{010} &  C \\
    0 & C & P^{100} 
    \end{bmatrix}\xrightarrow{\begin{bmatrix}0 & 0 &  1 \\0 & 0 &  0 \\1 & 0 & 0\end{bmatrix}}
    \begin{bmatrix}
    P^{001} & C &  0 \\
    C & P^{010} &  0 \\
    0 & 0 & P^{100} 
    \end{bmatrix}\\
    &\xrightarrow{\frac{\sqrt{2}}{2}\begin{bmatrix}1 & -1 &  0 \\1 & 1 &  0 \\0 & 0 & 0\end{bmatrix}}
    \begin{bmatrix}
    P^{001}-C & 0 &  0 \\
    0 & P^{010}+C &  0 \\
    0 & 0 & P^{100} 
    \end{bmatrix}.
\end{aligned}
\end{equation}
From this reduced density matrix, one can see that the population in the excited level of the second spin increases while that of the mediator decreases after the cascade interaction. Since the population relates directly to the energy, this demonstrates that heat transport is realized through coherence. Furthermore, the amount of heat transferred is proportional to the coherence strength.

\section{Coherence consumption and heat transfer}
\begin{figure}[htbp]
    \centering
    \includegraphics[width=1\linewidth]{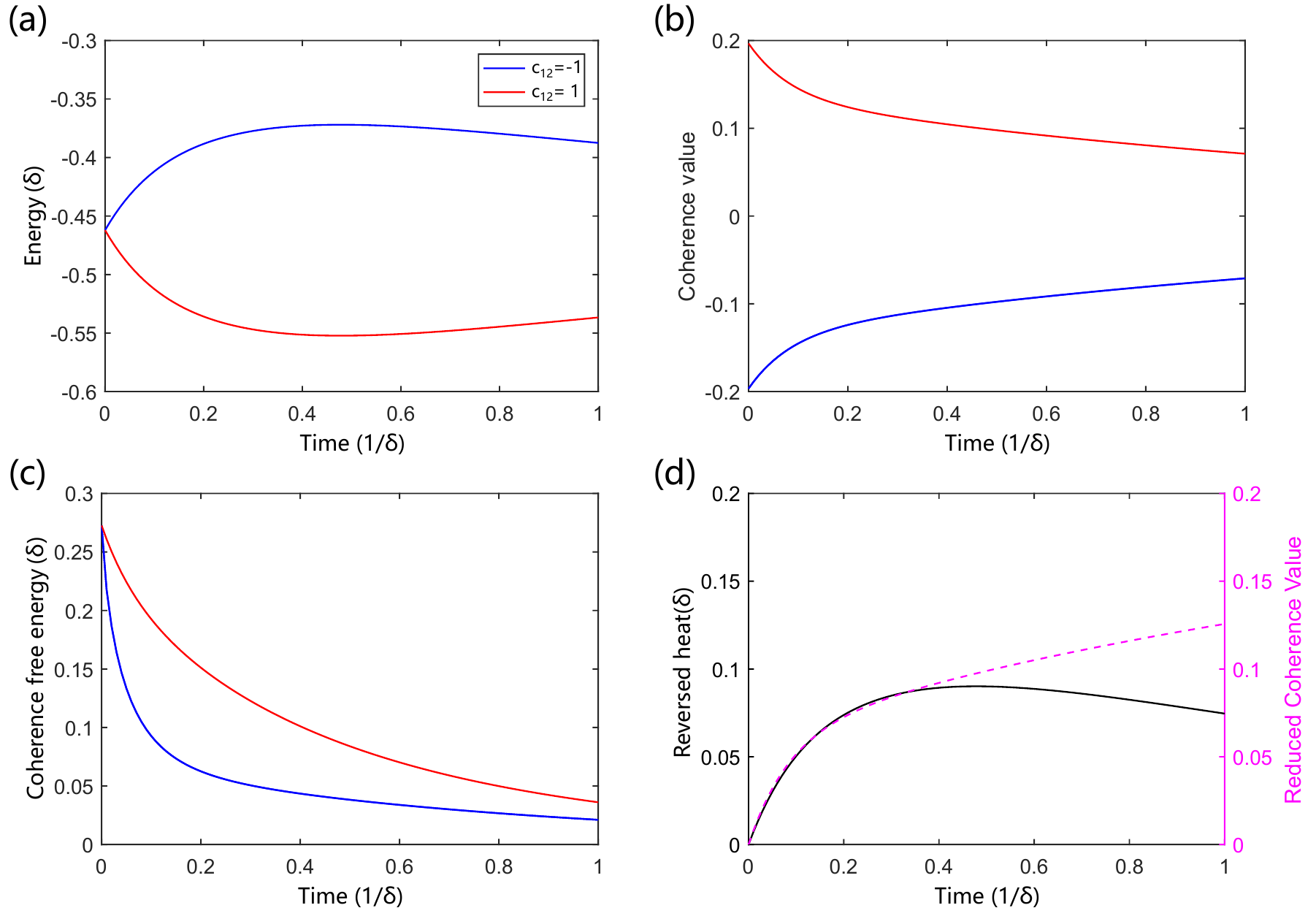}
    \caption{Energy and coherence changes in collision model with cascade interaction.}
    \label{EaC}
\end{figure}

We investigate the relationship between coherence consumption and heat transfer using a two-subsystem collision model. The total Hamiltonian is given by $\mathrm{H}_{t2} = \frac{\delta}{2} (\sigma_{z}^1 + \sigma_{z}^2 + \sigma_{z}^r)$, and the local temperature of each spin is set to $\delta / k_B$. Thermal contact is implemented via the unitary operation described in the main text, with a coupling strength of $g = 20\delta$ and a contact time per operation of $t_c = 0.01$. Simulation results for different initial coherence values $c_{12}$ are presented in Figure~\ref{EaC}.

From these results, we conclude that heat can flow against the temperature gradient when coherence is consumed. This is evidenced by the simultaneous decrease in both the absolute coherence magnitude and the coherence free energy, while the amount of reversed heat flow increases. At the onset of the cascade contact, changes in coherence and energy are closely correlated.

\section{Phase of coherence}
Coherence is characterized by two parameters: its strength and its phase. We have shown that the strength determines how much energy can be modulated. Here, we illustrate the physical meaning of the phase, which dictates the tendency and direction of energy changes under specific operations. We use the same three models as before to demonstrate this effect.

For the single-spin model, the function of coherence can be understood using the Bloch sphere representation. The density matrix of a state with zero energy and arbitrary coherence is
\begin{equation}
    \rho_1^\mathrm{ph}=
    \begin{bmatrix}
    1/2 & e^{i\alpha}\lambda\\
    e^{-i\alpha}\lambda & 1/2
    \end{bmatrix}=
    \begin{bmatrix}
    1/2 & (\cos{\alpha}+i\sin{\alpha})\lambda\\
    (\cos{\alpha}-i\sin{\alpha})\lambda & 1/2
    \end{bmatrix}=\frac{1}{2}I_2+\lambda(\cos{\alpha}\sigma_x-\sin{\alpha}\sigma_y),
\end{equation}
where $I_2$ is the $2 \times 2$ identity matrix. This corresponds to a vector on the Bloch sphere. For the Hamiltonian $\mathrm{H}_1 = \delta\sigma_z / 2$, the energy is determined by the z-axis component, while the coherence represents the projection of this vector onto the xy-plane: the real part is its projection onto the x-axis, and the imaginary part onto the y-axis.

Unitary operations act as rotations of this vector. Here, we focus on operations generated by operators like $\sigma_x$, which correspond to rotations around the x-axis. This is because, for exchanges of populations between two levels in higher-dimensional systems, the exchange operations are usually generated by real couplings such as $\sigma_-^p\sigma_+^q+\sigma_+^p\sigma_-^q$ (where $\sigma_+$ and $\sigma_-$ are the creation and annihilation operators). In this single-spin case, when the Bloch vector rotates about the x-axis, the x-component remains unchanged, and only the y-component of the coherence influences the energy, which is determined by the z-component:
\begin{equation}
    \frac{I_2+\sigma_y}{2}=
    \begin{bmatrix}
    1/2 & -i/2\\
    i/2 & 1/2
    \end{bmatrix}\xrightarrow{e^{-i\sigma_x\pi/4}}
    \begin{bmatrix}
    1 & 0\\
    0 & 0
    \end{bmatrix}=\frac{I_2+\sigma_z}{2}.
\end{equation}

A positive y-component converts to a positive z-component under an infinitesimal rotation. The phase of the y-component in the upper right element of the density matrix is $3\pi/2$. Thus, it is reasonable to define the efficiency of coherence under such an operation as its projection onto the y-axis:
\begin{equation}
    \eta=\cos(\alpha+3\pi/2)=\cos(\alpha+\pi/2).
\end{equation}

This definition can be verified using the second model. Coherence with zero phase does not affect the energy under normal coupling:
\begin{equation}
    \begin{bmatrix}
    0 & 0 & 0 & 0 \\
    0 & 1/2 & 1/2 & 0 \\
    0 & 1/2 & 1/2 & 0 \\
    0 & 0 & 0 & 0
    \end{bmatrix}\xrightarrow{e^{-i(\sigma_-^1\sigma_+^2+\sigma_+^1\sigma_-^2)\pi/4}}
    \begin{bmatrix}
    0 & 0 & 0 & 0 \\
    0 & 1/2 & 1/2 & 0 \\
    0 & 1/2 & 1/2 & 0 \\
    0 & 0 & 0 & 0
    \end{bmatrix}.
\end{equation}
In contrast, coherence with phase $\alpha = 3\pi/2$ causes heat to transfer from the first spin to the second:
\begin{equation}
    \begin{bmatrix}
    0 & 0 & 0 & 0 \\
    0 & 1/2 & -i/2 & 0 \\
    0 & i/2 & 1/2 & 0 \\
    0 & 0 & 0 & 0
    \end{bmatrix}\xrightarrow{e^{-i(\sigma_-^1\sigma_+^2+\sigma_+^1\sigma_-^2)\pi/4}}
    \begin{bmatrix}
    0 & 0 & 0 & 0 \\
    0 & 1 & 0 & 0 \\
    0 & 0 & 0 & 0 \\
    0 & 0 & 0 & 0
    \end{bmatrix}.
\end{equation}

The situation is more complex in the third model because the cascade interaction involves multiple sequential interactions, and all contributions must be considered. Here, we analyze how coherence changes in the first contact and how this result affects the second contact in the cascade. Both contacts evolve under the coupling Hamiltonian described in the main text. In the subspace spanned by the basis states $\left|0_1 0_2 1_m\right\rangle$, $\left|0_1 1_2 0_m\right\rangle$, and $\left|1_1 0_2 0_m\right\rangle$, the reduced density matrix evolves as:
\begin{equation}
\begin{aligned}
    &\begin{bmatrix}
    P^{001} & 0 &  0 \\
    0 & P^{010} &  e^{i\alpha}\lambda \\
    0 & e^{-i\alpha}\lambda & P^{100} 
    \end{bmatrix}\xrightarrow{e^{-i(\sigma_-^1\sigma_+^m+\sigma_+^1\sigma_-^m)\theta}}\\
    &\begin{bmatrix}
    P^{001} & -ie^{-i\alpha}\lambda\sin\theta &  0 \\
    ie^{i\alpha}\lambda\sin\theta & P^{010} &  e^{i\alpha}\lambda\cos\theta \\
    0 & e^{-i\alpha}\lambda\cos\theta & P^{100} 
    \end{bmatrix}\xrightarrow{e^{-i(\sigma_-^2\sigma_+^m+\sigma_+^2\sigma_-^m)\theta}}\\
    &\begin{bmatrix}
    P^{001}+\lambda\cos\alpha\sin\theta\sin{2\theta} & -i\lambda\sin\theta(e^{-i\alpha}\cos^2\theta-e^{i\alpha}\sin^2\theta) &  -ie^{i\alpha}\lambda\sin\theta\cos\theta \\
    i\lambda\sin\theta(e^{i\alpha}\cos^2\theta-e^{-i\alpha}\sin^2\theta) & P^{010}-\lambda\cos\alpha\sin\theta\sin{2\theta} &  e^{i\alpha}\lambda\cos^2\theta \\
    ie^{-i\alpha}\lambda\sin\theta\cos\theta & e^{-i\alpha}\lambda\cos^2\theta & P^{100} 
    \end{bmatrix}.
\end{aligned}
\end{equation}

In the first contact, the coherence between the two subsystems is converted to coherence between the second subsystem and the mediator, with the new coherence phase being $-\alpha + 3\pi/2$. Therefore, the efficiency of the coherence in the second contact is given by $\eta = \cos(-\alpha + 3\pi/2 + \pi/2) = \cos\alpha$. As a result, the population of $\left|0_1 0_2 1_m\right\rangle$ increases while that of $\left|0_1 1_2 0_m\right\rangle$ decreases by the same amount $\lambda \cos\alpha \sin\theta \sin(2\theta)$ after the cascade interaction, indicating that heat $\lambda \cos\alpha \sin\theta \sin(2\theta)$ flows from the second subsystem to the mediator. Thus, $\cos\alpha$ indeed quantifies the phase-dependent efficiency of the coherence.

\section{Contact at different temperatures}
All the previous analyses are based on the assumption that the local subsystem and the mediator are at the same temperature. To examine how coherence affects heat flow when they have different temperatures, we analyze the subspace spanned by the basis states $\left|0_1 0_2 1_m\right\rangle$, $\left|0_1 1_2 0_m\right\rangle$, and $\left|1_1 0_2 0_m\right\rangle$ in the two-subsystem cascade model. For a more precise condition under which the natural heat flow direction can be reversed, both paths linking the coherence levels must be taken into account.

The evolution of the reduced density matrix in the subspace of the first path ($\left|1_1 0_2 0_m\right\rangle$, $\left|0_1 0_2 1_m\right\rangle$, and $\left|0_1 1_2 0_m\right\rangle$) during the first cascade contact is:
\begin{equation}
 \begin{aligned}
    &\begin{bmatrix}
    P^{001} & 0 &  0 \\
    0 & P^{010} &  e^{i\alpha}\lambda_0 \\
    0 & e^{-i\alpha}\lambda_0 & P^{100} 
    \end{bmatrix}\xrightarrow{e^{-i(\sigma_-^1\sigma_+^m+\sigma_+^1\sigma_-^m)\theta}}\\
    &\begin{bmatrix}
   \cos^2\theta P^{001}+sin^2\theta P^{100} & -i\sin\theta e^{-i\alpha}\lambda_0 &  \cos\theta\sin\theta(iP^{100}-iP^{001}) \\
    i\sin\theta e^{i\alpha}\lambda_0 & P^{010} &  \cos\theta e^{i\alpha}\lambda_0 \\
    \cos\theta\sin\theta(iP^{001}-iP^{100}) & \cos\theta e^{-i\alpha}\lambda_0 & \cos^2\theta P^{100}+sin^2\theta P^{001} 
    \end{bmatrix}.
\end{aligned}   
\end{equation}
Similarly, for the second path:
\begin{equation}
 \begin{aligned}
    &\begin{bmatrix}
    P^{011} & e^{i\alpha}\lambda_1 &  0 \\
    e^{-i\alpha}\lambda_1 & P^{101} &  0 \\
    0 & 0 & P^{110} 
    \end{bmatrix}\xrightarrow{e^{-i(\sigma_-^1\sigma_+^m+\sigma_+^1\sigma_-^m)\theta}}\\
    &\begin{bmatrix}
    \cos^2\theta P^{011}+sin^2\theta P^{110} & \cos\theta e^{i\alpha}\lambda_1 &  \cos\theta\sin\theta(iP^{011}-iP^{110}) \\
    \cos\theta e^{-i\alpha}\lambda_1 & P^{101} &  i\sin\theta e^{-i\alpha}\lambda_1 \\
    \cos\theta\sin\theta(iP^{110}-iP^{011}) & -i\sin\theta e^{i\alpha}\lambda_1 & \cos^2\theta P^{110}+sin^2\theta P^{011} 
    \end{bmatrix}.
\end{aligned}   
\end{equation}
Here, $\lambda_{0(1)} = P^{0(1)}_m \lambda$ denotes the coherence distributed along each path, where $P^{0(1)}_m$ is the ground (excited) state population of the mediator. 

To evaluate the energy changes during the second cascade contact, we focus on levels $\left|0_1 0_2 1_m\right\rangle$ and $\left|0_1 1_2 0_m\right\rangle$ in the first matrix and $\left|1_1 0_2 1_m\right\rangle$ and $\left|1_1 1_2 0_m\right\rangle$ in the second. The populations of the remaining levels remain unaffected by this step. These yield two effective $2 \times 2$ reduced density matrices:
\begin{equation}
    \begin{bmatrix}
   \cos^2\theta P^{001}+sin^2\theta P^{100} & -i\sin\theta e^{-i\alpha}\lambda_0 \\
    i\sin\theta e^{i\alpha}\lambda_0 & P^{010} & 
    \end{bmatrix}, \text{and} 
    \begin{bmatrix}
     P^{101} &  i\sin\theta e^{-i\alpha}\lambda_1 \\
    -i\sin\theta e^{i\alpha}\lambda_1 & \cos^2\theta P^{110}+sin^2\theta P^{011} 
    \end{bmatrix}.
\end{equation}
Tracing out the first subsystem, the corresponding elements must be summed, since the bases represent the same physical states for the last two spins. This yields a combined $2 \times 2$ matrix:
\begin{equation}
    \begin{bmatrix}
   \cos^2\theta P^{001}+sin^2\theta P^{100}+P^{101} & -i\sin\theta e^{-i\alpha}(\lambda_0- \lambda_1) \\
    i\sin\theta e^{i\alpha}(\lambda_0- \lambda_1) & P^{010}+\cos^2\theta P^{110}+sin^2\theta P^{011} 
    \end{bmatrix}.
\end{equation}
This matrix behaves like the density matrix of an effective two-level system under the second cascade contact, which is equivalent to a unitary operation $e^{-i \sigma_x \theta}$ acting on it. Consequently, the condition for heat flow reversal can be interpreted by viewing this matrix as representing an effective qubit subject to a rotation.

When the bath temperature is higher than that of the system, the inequality $\cos^2\theta P^{001}+sin^2\theta P^{100}+P^{101}>P^{010}+\cos^2\theta P^{110}+sin^2\theta P^{011}$ holds naturally, and reversal means this difference grows further. This occurs if and only if the z-component does not decrease under the unitary, leading to the condition:
\begin{equation}
    2\sin\theta \cos\alpha (\lambda_0- \lambda_1)>\tan\theta[(\cos^2\theta P^{001}+sin^2\theta P^{100}+P^{101})-(P^{010}+\cos^2\theta P^{110}+sin^2\theta P^{011})].
\end{equation}
Similarly, when the bath temperature is lower than the system's, the condition for reversal is:
\begin{equation}
    2\sin\theta \cos\alpha (\lambda_1- \lambda_0)>\tan\theta[(P^{010}+\cos^2\theta P^{110}+sin^2\theta P^{011})-(\cos^2\theta P^{001}+sin^2\theta P^{100}+P^{101})].
\end{equation}

Taking the limit $\theta \to 0$ (equivalently, the contact time approaches zero), these conditions simplify to: 
\begin{equation}
\begin{aligned}
    &2\sin\theta \cos\alpha (\lambda_0- \lambda_1)>\tan\theta[(\cos^2\theta P^{001}+sin^2\theta P^{100}+P^{101})-(P^{010}+\cos^2\theta P^{110}+sin^2\theta P^{011})]\\
    &\to 2\cos\alpha (\lambda_0- \lambda_1)>(P^{001}+P^{101})-(P^{010}+ P^{110})\\
    &\to 2\lambda \cos\alpha(P^0_m-P^1_m)>P^0_sP^1_m-P^1_sP^0_m\\
    &\to  2\lambda\cos\alpha>\frac{e^{\beta_s\delta}-e^{\beta_m\delta}}{(e^{\beta_s\delta}+1)(e^{\beta_m\delta}-1)},
\end{aligned}
\end{equation}
and
\begin{equation}
\begin{aligned}
    &2\sin\theta \cos\alpha (\lambda_1- \lambda_0)>\tan\theta[(P^{010}+\cos^2\theta P^{110}+sin^2\theta P^{011})-(\cos^2\theta P^{001}+sin^2\theta P^{100}+P^{101})]\\
    &\to  2\lambda\cos\alpha<\frac{e^{\beta_s\delta}-e^{\beta_m\delta}}{(e^{\beta_s\delta}+1)(e^{\beta_m\delta}-1)}.
\end{aligned}
\end{equation}
Here, $P^{0(1)}_s$ denotes the ground (excited) state population of the subsystem. These match the reversal conditions derived from the apparent temperature analysis.

\end{document}